\documentclass[useAMS,usenatbib]{mn2e}
\usepackage{graphicx}
\usepackage{latexsym}
\usepackage{psfig}
\usepackage{amssymb,amsmath}
\usepackage{subfig}
\usepackage{caption}
\usepackage{supertabular}
\usepackage{longtable}
\usepackage{ltcaption}

\title{The dangers of being trigger--happy}

\author[J. E. Dale,T. J. Haworth,E. Bressert]{J. E. Dale$^{1,2}$\thanks{E-mail: dale@usm.lmu.de (JED)}, T. J. Haworth$^{3}$, E. Bressert$^{4}$\\
$^{1}$Excellence Cluster `Universe', Boltzmannstr. 2, 85748 Garching, Germany.\\
$^{2}$Universit\"{a}ts Sternwarte M\"{u}nchen, Scheinerstr. 1, 81679 M\"{u}nchen, Germany.\\
$^{3}$Institute of Astronomy, Madingley Road, Cambridge, CB3 0HA, UK.\\
$^{4}$European Southern Observatory, Karl--Schwarzschild--Str. 2, 85748 Garching, Germany.}

\begin{document}

\pagerange{\pageref{firstpage}--\pageref{lastpage}} \pubyear{2006}

\maketitle

\label{firstpage}

\def\mnras{MNRAS}
\def\apj{ApJ}
\def\aj{AJ}
\def\aap{A\&A}
\def\apjl{ApJL}
\def\apjs{ApJS}
\def\araa{ARA\&A}
\def\pasj{PASJ}
 
\begin{abstract}
We examine the evidence offered for triggered star formation against the backdrop provided by recent numerical simulations of feedback from massive stars at or below giant molecular cloud sizescales. We compile a catalogue of sixty--seven observational papers, mostly published over the last decade, and examine the signposts most commonly used to infer the presence of triggered star formation. We then determine how well these signposts perform in a recent suite of hydrodynamic simulations of star formation including feedback from O--type stars performed by Dale et al (2012a, b, 2013a, b, 2014). We find that none of the observational markers improve the chances of correctly identifying a given star as triggered by more than factors of two at most. This limits the fidelity of these techniques in interpreting star formation histories. We therefore urge caution in interpreting observations of star formation near feedback--driven structures in terms of triggering.
\end{abstract}

\begin{keywords}
stars: formation
\end{keywords}
 
\section{Introduction}
Star formation is a pivotal process in astrophysics, impacting fields from galactic evolution and possibly cosmic reionisation at the very largest scales to the formation of planets and life at the smallest. A crucial part of the process is the feedback exerted by the stars themselves, which has long been recognized to have two opposing aspects. Expanding bubbles driven by photoionisation, winds and supernova explosions can accelerate cold gas to beyond the escape velocities of GMCs, potentially destroying them and shutting star formation down. This is generally termed `negative feedback' and has often been blamed for the long--standing problem of the low star formation efficiencies or slow star formation rates in molecular clouds \citep{1979IAUS...84...35S}.\\
\indent However, the gas density in swept--up shells can be high enough that the shells become gravitationally unstable and form additional stars. This process is usually referred to as `triggered' star formation or positive feedback, and is a popular subject for observers and theorists alike. As we discuss below, there are several other processes acting at larger scales, which can also be thought of as triggering star formation. However, they have not attracted as much attention as feedback--triggered star formation, perhaps because the latter raises the intriguing possibility that star--formation may be a self--propagating phenomenon \citep{1981ApJ...249...93S}.\\
\indent Simulations of star formation including the effects of stellar feedback are now possible thanks to improvements in algorithms to include new physics e.g. \cite{2005MNRAS.358..291D,2006ApJ...640L.187L,2007ApJ...656..959K,2010ApJ...711.1017P,2011ApJ...729...72P,2011ApJ...740...74K,2011ApJ...736..142B,2012A&A...546A..33T,2012MNRAS.420..562H,2012MNRAS.426..203H,2013MNRAS.435..917W,2013MNRAS.431.3470H,2014MNRAS.tmp..421M}. However, the tension between the negative and positive aspects of stellar feedback is still a subject of debate.\\
\indent In this paper, we will concentrate on the positive aspects of feedback from stars, namely triggered star formation on scales of molecular clouds. In Section 2, we attempt to define triggering more carefully. In Section 3, we differentiate between several different lengthscales at which triggered star formation could be imagined to occur to separate stellar feedback internal to GMCs from other processes. In Section 4, we discuss some of the very large body of observational studies of triggered star formation and examine the different criteria used by various authors to infer triggering. In Section 5 we discuss hydrodynamical simulations relating to triggering. Our discussion and conclusions follow in Section 6 and 7. The catalogue of observation papers we have examined are included in an Appendix, in a checkbox table showing which triggering signposts are used in each publication.\\
\section{Definitions of triggering}
Before beginning, it is helpful to expend some effort in defining what is actually meant by \emph{triggered star formation}, since it does not have a single universally--agreed--upon definition, as discussed in \cite{2007MNRAS.377..535D} and \cite{2013MNRAS.430..234D}. It could be taken to mean one or more of the following things, which we denote Type I, II or III triggering:
\begin{itemize} 
\item {\bf Type I triggering}: A temporary or long--term increase in the star formation rate.
\item {\bf Type II triggering}: An increase in the final star formation efficiency.
\item{\bf Type III triggering}: An increase in the total final number of stars formed.
\end{itemize}
\indent A moment's reflection will show that these definitions are \emph{not} equivalent. For example, the action of feedback could be to bias the initial mass function slightly towards lower stellar masses, increasing the numbers of stars but not necessarily increasing the rate at which gas is consumed, or the final stellar mass. Alternatively, an expanding HII region may cause a dense clump to collapse earlier and faster than it would have otherwise, thus increasing the star formation rate there, but not necessarily changing either the final star formation efficiency or number of stars. \cite{2007MNRAS.377..535D} referred to such a case as `weak triggering'. Over longer timescales or large distance scales, only Type II and III triggering -- the forcing of a system to create a larger stellar mass or more stellar objects -- are of interest we refer to these as `strong' triggering.\\
\indent One could potentially add a fourth type of triggering in which some statistical property of the stars formed is different, for example the initial mass function. Some theoretical work \citep[e.g][]{1994A&A...290..421W} predict that triggered star formation may result in the production of an excess of massive stars. However, there is no convincing observational evidence as yet for variations in the initial mass function, and \cite{2013MNRAS.430..234D} and \cite{2013MNRAS.431.1062D} found only modest variations in the IMF caused by ionisation feedback.\\
\indent An additional aspect to consider is that these definitions can all be local or global. Simulations by \cite{2012MNRAS.427.2852D,2013MNRAS.431.1062D} found that the overall effect of ionisation feedback on molecular clouds scales was almost always negative in the sense of lowering the star formation efficiency over a given timescale and resulting in fewer stars being born, so that neither Type II or III triggering occurred \emph{globally}. However, by tracing the histories of the gas from which stars formed, they were able to establish the occurrence of localised Type III strong triggering, where stars that would not otherwise form were induced to do so by feedback. These results are reconciled by the fact that many stars which would form in the absence of feedback are aborted by the destruction of much of the densest star--forming gas by the O--stars.\\
\indent However, these findings were only possible through comparison of pairs simulations started from identical initial conditions and run with and without feedback. For any of the above definitions to be meaningful, it must be possible to identify a baseline rate, efficiency or number of stars against which the putative increase can be measured. Type I triggering could be inferred with high confidence if the star formation history of a given system could be measured and one or more sudden jumps in the star formation rate could be inferred that could be strongly connected to a likely triggering event, such as if a jump coincided with the birth of massive stars. This could in principle be inferred from observing a system at a single given time if the ages of all the stars could be measured very accurately. This is very difficult in practice, and in any case relies on the assumption that the triggering agent is able to abruptly and substantially increase the star formation rate, which theoretical work shows is unlikely to be true.\\
\indent Type II and III triggering can only be confidently inferred if a credible model can be proposed of how much stellar mass or how many stars the system in question would have formed if the triggering process had not been operating. While an easy task for simulators, this approach is extremely difficult from an observational point of view and is rarely attempted.\\
\indent As detailed in Section 4, in most cases what is actually measured by observers is the spatial association of young stellar objects with a feedback--driven structure such as a shell or a bright--rimmed cloud, and this is taken to be indicative of triggering. Since it is often very difficult to compute the masses or ages of YSO's, the implicit definition of triggering most commonly used is an increase in the total number of stars, i.e. Type III. However, as we say above, in order to be confident that such geometrical correlations show triggering, a model of what the system would look like, and how many stars it would contain, in the absence of the triggering agent is strictly required.\\
\section{Scales of triggering}
Star formation is a pervasive process which can be studied on scales ranging from individual stars and proto--planetary disks (tens of AU) up to whole galaxies (tens or even a few hundred kpc). There are a concomitant range of possible triggering mechanisms acting over this wide range of lengthscales.\\
\indent Collisions of galaxies resulting in the overpressuring of their ISM and inducing enormous star formation rates are generally referred to as starbursts and are often observed as ultra luminous infrared galaxies (ULIRGs, \cite{2012ARA&A..50..531K}). Such events would likely satisfy the conditions of Type I, II and III triggering  simultaneously.\\
\indent Supershells are very large-scale ($>$100pc) structures driven by the combined feedback from a rich cluster or OB--association. These structures are much larger than the parent GMC of the driving cluster and are able to sweep up large quantities of the intracloud ISM, leading to gravitational instability and eventually star formation (see \cite{1988ARA&A..26..145T} for a review).\\
\indent In the previous two processes, it is not easy to disentangle triggered molecular \emph{cloud} formation from triggered \emph{star} formation. Given that very few starless molecular clouds are known, the former seems to lead inexorably to the latter. \cite{2011ApJ...741...85D} present combined HI and $^{12}$CO images of two objects morphologically midway between shells and chimneys. They find numerous molecular clouds embedded in the shells' walls, often many times further from the Galactic plane than the scale--height of molecular material. This makes it unlikely that the clouds were pre--existing objects or have been transported in molecular form from the plane. Instead it is much more probable that the formation of the molecular gas has been triggered \emph{in situ} by the expansion of the shells/chimneys. They confirm that many of these apparently triggered clouds are indeed forming stars.\\
\indent Using the abrupt initiation and completion of star formation and the existence of a nearby OB association whose O--stars should have exploded at the right epoch and distance,  \cite{2001ASPC..243..791P} concluded that star formation in the Sco--Cen OB association may have been triggered by supernovae striking a molecular cloud. This would likely be a case of Type I triggering, as it is probable that the cloud would eventually have started to collapse of its own accord.\\
\indent Another scenario in which star formation is at least accelerated by an interaction involving pre--existing molecular clouds is the collision of GMCs \citep{1998ASPC..148..150E}. \cite{2000ApJ...536..173T} propose that the global star formation rate in disk galaxies is regulated by the collision of clouds. In this model, \emph{all} star formation is triggered by this process, in which case the word `triggered' strictly loses its explanatory power. Cloud--cloud collisions would then just be another link in the star formation process, having the same status as gravity itself.\\
\indent Whether or not the star formation process is overall regulated by cloud--cloud collisions, there is some evidence of such interactions in progress. \cite{2009ApJ...696L.115F} and \cite{2010ApJ...709..975O} present combined molecular and infrared observations of the Westerlund 2 super star cluster, its associated RCW49 HII region and the molecular gas in the surrounding few tens of pc. They find strong evidence for two roughly equal--mass converging velocity components in the molecular gas separated by $\approx15$kms$^{-1}$ and centred on Wd 2, inferring that the formation of the cluster was triggered by a collision. \cite{2011ApJ...738...46T} and  \cite{2014ApJ...780...36F} use similar methods and reach similar conclusions concerning M20 and NGC 3603 respectively.\\
\indent The scenarios mentioned above can be grouped together under the heading \emph{external triggering}, since they all invoke some agent external to a given molecular cloud causing it to begin forming stars at a given epoch and possibly leading it to form more stars than it would if it were left alone.\\
\indent The final possibility, and the most popular in terms of the quantity of literature devoted to it, is \emph{internal triggering}. This denotes triggering due to feedback from previously--formed stars \emph{embedded within} a given cloud. This is generally the mechanism acting on the smallest scales of a few pc, but this form of triggering may act on the scale of whole clouds up to $\sim100$pc.\\
\indent Stellar feedback (jets/outflows, photoionisation, winds, supernovae) can act positively or negatively from the point of view of star formation. By definition it can only influence clouds in which star formation is already underway and containing a rich density and velocity field. Overdensities in the cloud have the potential to form stars, but can also be disrupted by turbulence. Local increases in the gas or dynamic pressure could precipitate otherwise stable or transient cores to collapse. This process is usually referred to as radiation--driven implosion or cloud--crushing \citep[e.g.][]{1994A&A...289..559L,1994ApJ...420..213K}. Even smooth and diffuse gas can be induced to form stars if collected together rapidly enough, as in the collect--and--collapse process \cite[e.g.][]{1977ApJ...214..725E,1994A&A...290..421W}. It is these two process or variants of them, which have attracted most attention in the study of triggering on scales of individual molecular clouds or smaller.\\
\section{Observations of internal triggering}
\begin{figure}
\includegraphics[width=0.50\textwidth]{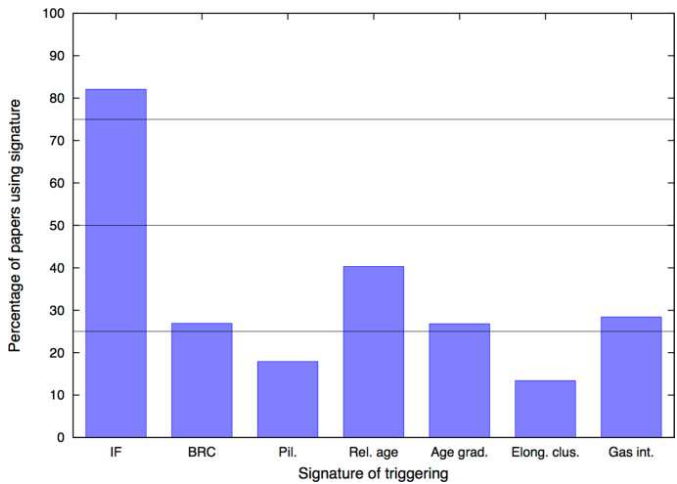}
\caption{Barchart illustrating what fraction of all the papers listed in Table \ref{tab:bigtable}  cite each of the seven triggering signposts.}
\label{fig:barchart}
\end{figure}

\indent Internal triggering by feedback from massive stars has received a great deal of observational attention, particularly recently with the advent of space telescopes such as \emph{Spitzer}, \emph{Herschel} and \emph{WISE} that are able to peer deep into star--forming regions in great detail. We surveyed a total of 71 papers published between the years of 1984 and 2015, mostly during the last decade. This is not an exhaustive list of every paper published on this topic between these years but, we believe, covers a representative cross section involving several different methods of inferring triggering.\\
\indent In Table \ref{tab:bigtable} in the Appendix, we list the papers arranged alphabetically by author (first column). We omit four papers which did not find any evidence of triggering in their observed targets, namely \cite{2009A&A...504...97B} (observing IC1396), \cite{2013MNRAS.429.1386D} (observing N14) \cite{2013A&A...550A.116T} (observing the IRDC G18.93-0.03) and \cite{2013A&A...557A..29C} (observing the Rosette). We should also point out that many authors, e.g. \cite{2009A&A...497..789U,2014A&A...566A.122S,2015ApJ...798...30L}, urge caution and do not present their results as \emph{conclusive} evidence for triggered star formation.\\ 
\indent The second column briefly describes the target(s) of the paper. The following seven checkbox columns detail the most common signposts used by the authors to infer that a given stellar population was partially or wholly triggered: (1) Shell/IF/HIIR: proximity to feedback--driven shells, ionisation fronts or HII regions (2) BRC: proximity to a bright--rimmed cloud, (3) Pil./Com.: proximity to a pillar or cometary cloud, (4) Re./Dyn. ages: use of relative ages of stars and dynamical ages of feedback--driven structures, (5) Age grad.: use of a gradient in ages of young stars pointing towards a feedback source, (6) Elong. clus.: use of the elongation of a cluster of young stars towards a feedback source, (7) Gas int.: use of evidence of strong interaction between a feedback source and star--forming gas, for example evidence of shocks or thermal overpressure. The final column gives the type of feedback source or structure that the authors credit with triggering star formation.\\
\indent Note that some of these categories overlap and judgement about which have been invoked by a given author can have a subjective element depending on the language used. However, Table \ref{tab:bigtable} and Figure \ref{fig:barchart} still give an overview of how the community has inferred the presence of triggered star formation, in the sense of locally increasing the numbers of stars.\\
\indent Figure \ref{fig:barchart} shows a breakdown of what fraction of the papers listed in Table \ref{tab:bigtable} cite each of the seven triggering signposts, illustrating which of these mechanisms are the most popular. By far the most commonly--cited single indicator of triggering is the presence of young stars near a shell, ionisation front or HII region at 55/67 $\approx$82$\%$ of papers. Some authors, of course, say that the primary agent of triggering cannot be definitively identified, e.g. \cite{2006ApJ...649..759C,2007ApJ...670..428C}. These results are usually interpreted in the context of the collect--and--collapse model, although several authors, e.g. \cite{2006A&A...446..171Z,2007A&A...472..835Z} note that the fragment masses observed are often considerably higher than those predicted by theoretical models. Two other structural signposts that are commonly used to infer triggering are  bright--rimmed clouds (18/67 $\approx$27$\%$) and pillars or cometary clouds (12/67 $\approx$18$\%$). These are in general smaller features than shells, but are of course often to be found on shell perimeters.\\
\indent These first three categories are all essentially geometrical in nature -- they rely on placing the YSO's near some feature or structure of the ISM which is known to be caused by massive--star feedback.\\
\indent Most authors use more than one feedback indicator. The remaining four categories in the table are generally used to provide supporting evidence for triggering -- they are only very rarely invoked alone. The most often--used method (27/67 $\approx$40$\%$) is to check the ages of the YSO's relative to the feedback--generating stars or to the dynamical age of the feedback driven bubbles, or the crossing--times of bright--rimmed clouds. Age gradients in the YSO's pointing towards the massive stars are also commonly used to infer triggering by the gradual progression of an ionisation or shock front through a dense cloud (18/67 $\approx$27$\%$). Geometrically elongated clusters of young stars are invoked by 9/67 $\approx$13$\%$ of authors as evidence of the same process. Finally, direct evidence of strong interaction between dense molecular gas and an HII region or wind bubble is put forward by 19/67 $\approx$28$\%$ of authors. For example, \cite{2004A&A...414.1017T} and \cite{2009A&A...497..789U} only regard BRCs as strong candidates for hosting triggered star formation if the pressure in the ionised boundary layer is larger than or comparable to, the clouds' internal pressures. Detailed radiation--hydrodynamics calculations of model BRCs by \cite{2012MNRAS.426..203H,2013MNRAS.431.3470H} show that most of the observational diagnostics used, e.g. non--Gaussian line profiles, do accurately represent what is happening inside the clouds. Overall, 45/67 $\approx$67$\%$ of authors find some corroborating evidence beyond geometrical association to support the inference of triggering.\\
\indent Very few authors consider the counterfactual scenario which would obtain if the identified feedback source or feedback--driven structure were not present. One notable exception is \cite{2004A&A...414.1017T}, who explicitly show that several of their observed bright--rimmed clouds would be gravitationally stable if it were not for the pressure of the photoevaporation flows driven by the exciting HII regions. They therefore consider these objects strong candidates for radiation--driven implosion.\\
\indent \cite{2010ApJ...712..797B} state clearly that they regard an increase in star formation efficiency (SFE) as a good definition of triggering. They compute the star formation efficiency in the Vulpecula OB association. While they do see structures often associated with triggered star formation such as gaseous pillars, their measured star formation efficiency is consistent with those computed by \cite{2009ApJS..181..321E} and they say that they have no means of ascertaining whether the young stars would have formed in the absence of the nearby OB stars.\\
\indent Regarding what feedback agents or feedback--driven structures are cited by authors, HII regions are by the far the most popular at 53/67 ($\approx$79$\%$), with winds being invoked alone or in combination with ionisation 18/67 times ($\approx$27$\%$) and supernova remnants account for the remaining 5/67 $\approx$7$\%$. This may be related to the dominant role of HII regions as feedback agents, particularly at the early stages of GMC evolution before any O--stars have moved off the main--sequence, or it may reflect a bias that bubbles, which are actually composite structures, are often classified simply as HII regions.\\
\section{Hydrodynamic Simulations}
\indent A series of papers \citep{2012MNRAS.424..377D,2012MNRAS.427.2852D,2012MNRAS.424..377D,2013MNRAS.431.1062D,2013MNRAS.430..234D} have modelled the effects of internal and external photoionisation on star--forming clouds. The calculations were all performed with a Smoothed--Particle Hydrodynamics (SPH) code derived from that of \cite{1990nmns.work..269B}. Sink particles are used to model gravitational collapse and the formation of stars, as described in \cite{1995MNRAS.277..362B}. Photoionisation from O--type stars is modelled using a Str\"omgren--volume approach. Rays are drawn from radiation sources to gas particles and other particles intersecting the ray are used to compute the integrated recombination rate along the ray and hence to locate the ionisation front in that direction. Ionised particles are then heated to the canonical 10$^{4}$K. A detailed description of the algorithm can be found in \cite{2007MNRAS.382.1759D}. In cases where multiple ionising sources are present, an iteration over the sources is performed as described in \cite{2012MNRAS.427.2852D}. Other forms of feedback are neglected in these calculations. Several authors, \cite[e.g.][]{2002ApJ...566..302M}, have inferred semianalytically that photoionisation should be a more important source of feedback on GMC scales than winds, and it was shown in \cite{2014MNRAS.442..694D} that the addition of momentum--driven winds do indeed have only a very small effect on the outcome of the simulations presented here. None of our simulations are allowed to progress long enough for supernovae to begin exploding, so we do not do not discuss their effects and instead concentrate on triggering by photoionisation, which is also the main feedback mechanism cited in the observational works surveyed.\\
\indent The conclusions of these papers may be summarized as follows: (i) external photoionisation can produce modest increases in the SFR, SFE and numbers of stars, but only in clouds which are not vigorously forming stars already (ii) internal photoionisation always decreases the SFE and average SFR or leaves them unchanged, but may change the numbers of stars in either direction (iii) objects identified as triggered by comparison with feedback--free control runs are in general geometrically mixed with spontaneously--formed objects and thus the two populations are very hard to distinguish from a given single simulation snapshot. While it is true that local triggered star formation does occur in these calculations, it is therefore very difficult to identify triggered objects on a star--by--star basis simply by observing the end result of the simulations.\\
\indent We reinforce this point here by analysing the results of the Runs I, J, UF and UQ simulations presented in \cite{2012MNRAS.427.2852D,2012MNRAS.424..377D,2013MNRAS.431.1062D} and \cite{2013MNRAS.430..234D} in more detail and in terms of quantities which are easier to observe directly. The clouds were allowed to evolve from smooth initial states with imposed turbulent velocity fields, up to the point where each had formed three stars with a mass in excess of 20M$_{\odot}$. These stars, and any others which subsequently achieved this mass, were then regarded as ionising sources and feedback from them was enabled. Additionally, a control run without feedback was calculated for each simulation. Tracing of the particles from which stars formed in each [control, feedback] pair of simulations allowed which stars were triggered by feedback to be identified, as described in detail in \cite{2012MNRAS.424..377D}. The main properties of the clouds in the four simulations examined here are given in Table \ref{tab:clouds}.\\
\begin{table*}
\begin{tabular}{|l|l|l|l|l|l|l|l|l|l|}
Run&Mass (M$_{\odot}$)&R$_{0}$(pc)&$\langle n(H_{2})\rangle$ (cm$^{-3}$) & v$_{\rm RMS,0}$(km s$^{-1}$)&v$_{\rm RMS,i}$(km s$^{-1}$)&v$_{\rm esc,i}$(km s$^{-1}$)&$t_{\rm i}$ (Myr) &t$_{\rm ff,0}$ (Myr)\\
\hline
I&$10^{4}$&10&136&2.1&1.4&2.3&5.37&2.56\\
\hline
J&$10^{4}$&5&1135&3.0&1.8&3.5&2.09&0.90\\
\hline
UQ&$10^{4}$&5&1137&5.4&2.6&4.1&3.13&1.2\\
\hline
UF&$3\times10^{4}$&10&410&6.7&3.5&5.1&3.28&2.0\\
\hline
\end{tabular}
\caption{Initial properties of clouds listed in descending order by mass. Columns are the run name, cloud mass, initial radius, initial RMS turbulent velocity, RMS turbulent velocity at the time ionization becomes active, the escape velocity at the same epoch, the time at which ionization begins, and the initial cloud freefall time.}
\label{tab:clouds}
\end{table*}
\subsection{Association of triggered stars with shells, ionisation fronts and pillars}
\subsubsection{Shells}
\indent The most popular means of finding triggered star formation is looking for associations between between YSOs and shells, ionised bubbles or the ionisation fronts that bound them. The fragmentation of the shell driven by an over--pressured bubble expanding in a uniform medium is an appealingly--simple process. Such structures are readily identifiable observationally and accessible to analytic study \citep[e,g][]{1977ApJ...214..725E,1994A&A...290..421W,2001A&A...374..746W,2011ApJ...733...16I}. Analytic models have been compared with simulations with some degree of success \citep[e.g][]{2005ApJ...623..917H,2007MNRAS.375.1291D,2010MNRAS.407.1963W,2011ApJ...733...17I,2011MNRAS.411.2230D}. These simulations confirm the general concept of the collect--and--collapse model, but reveal that even such a simple system supports a rich phenomenology.\\
\indent No real molecular cloud is perfectly smooth and the complexity of this problem increases if the ambient gas is taken to be inhomogeneous. In a carefully--controlled series of simulations, \cite{2013MNRAS.435..917W} quantified the nature of the inhomogeneities in their background cloud using the fractal dimension of the medium. They found that smaller fractal dimensions results in smooth shells composed of a few very large clumps, resulting in clustered star formation. Larger fractal dimensions result in much less smooth shells, many pillar structures and more evenly distributed star formation. We note that all stars actually or potentially forming in any of the above--mentioned simulations are examples of Type II and III triggering, since the smooth or fractal clouds do not form any stars at all in the absence of feedback over the timescales considered.\\
\indent However, even in the case of triggering in a perfectly or relatively smooth cloud, there are issues which can complicate the identification of triggered stars. Projection effects ensure that some of the triggered stars appear to be inside the shell volume, where they may be confused with the stars belonging to the driving cluster. In addition, the triggered stars may not remain geometrically associated with the shell due their peculiar velocities which likely result in general in mixing of triggered objects with those spontaneously--formed. Apart from feedback, other processes such as large--scale turbulence, can create structures resembling shells which may well happen to have star formation in progress in the dense gas that defines their rims \citep{2011MNRAS.414..321D}.\\
\indent In simulations of turbulent clouds, it is often not possible to identify well--defined or well--cleared bubbles. However, even when it is, the stars found near the edges of the bubbles are found to be a mixture of triggered and spontaneous stars, due to the process of redistributed star formation discussed by \cite{2013MNRAS.431.1062D}.\\
\subsubsection{Ionisation fronts}
\begin{figure*}
\centering
\subfloat[Whole cloud]{\includegraphics[width=0.45\textwidth]{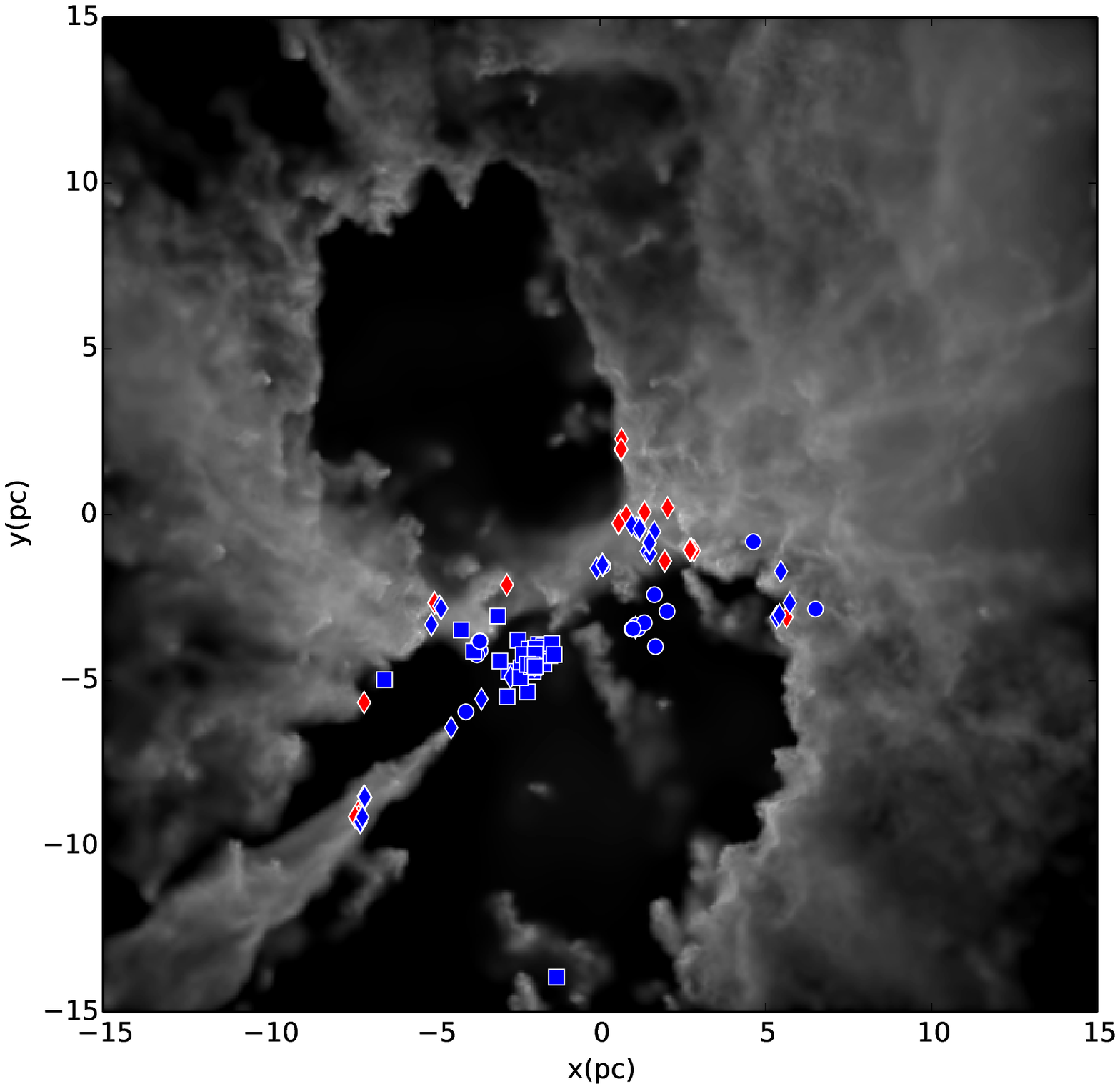}}
\hspace{0.1in}
\subfloat[Zoom on pillar]{\includegraphics[width=0.45\textwidth]{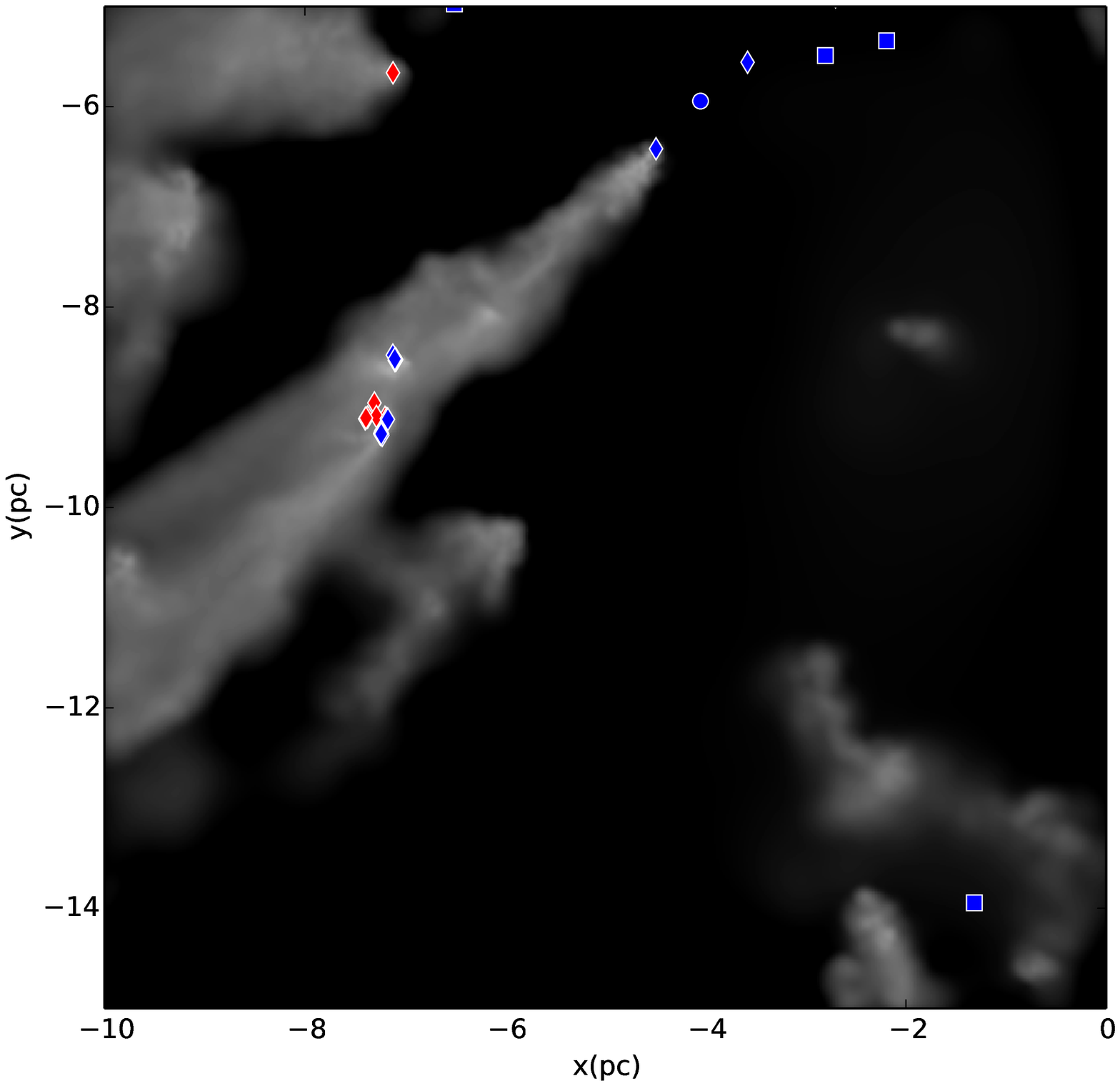}}
\caption{Relative positions of sink particles and gas in Run I showing the whole cloud in the left panel and a zoom on the prominent pillar in the lower left corner in the right panel. Sink particles are overlaid. Red sinks are triggered, blue sinks are spontaneously--formed, and symbol shapes denote sink ages: diamonds (age$<$1Myr); circles (1Myr$<$age$<$2Myr); squares (2Myr$<$age).}
\label{fig:runi_gas}
\end{figure*}

\begin{figure*}
\centering
\subfloat[Run I]{\includegraphics[width=0.45\textwidth]{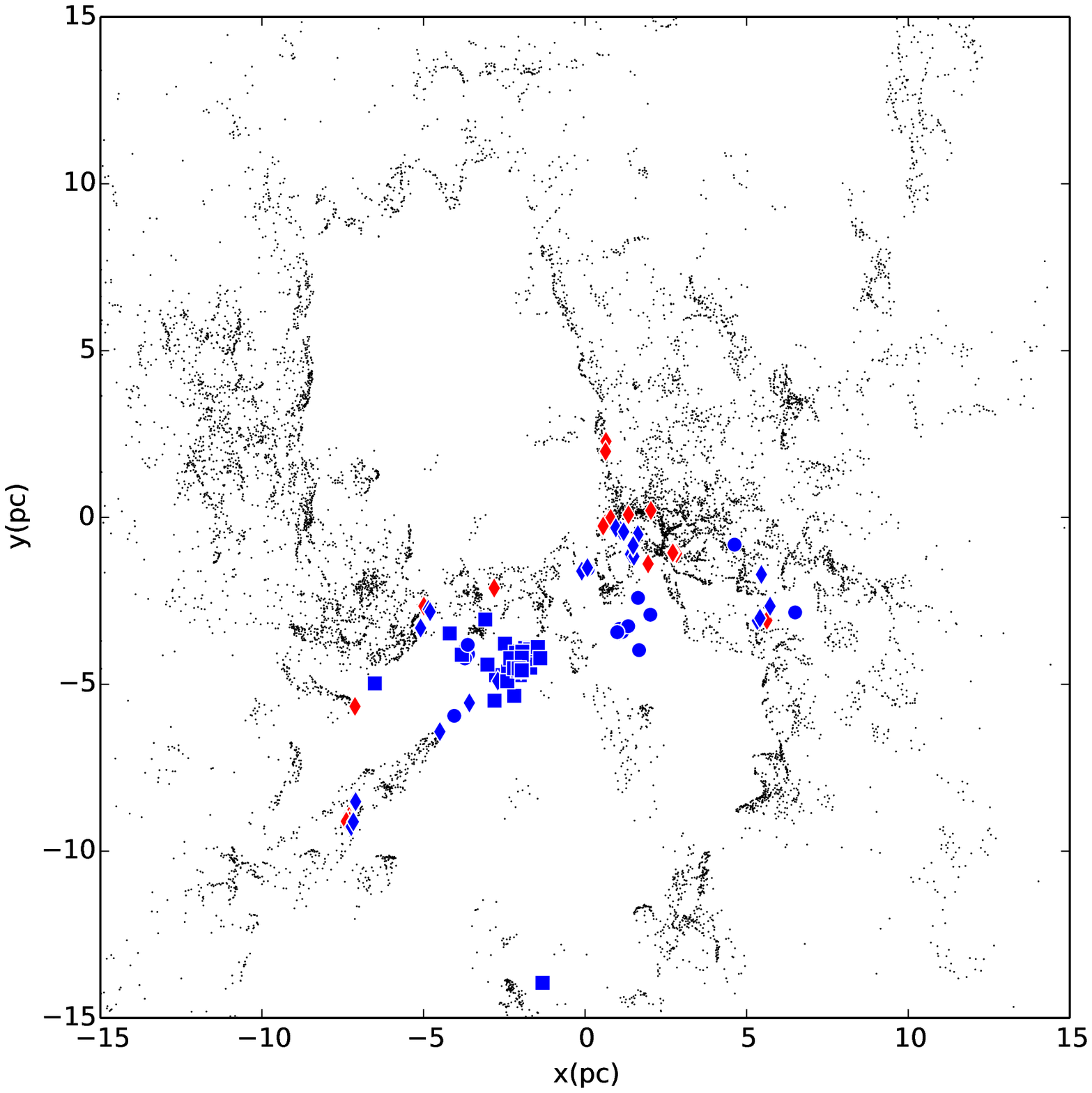}}
\hspace{0.1in}
\subfloat[Run J]{\includegraphics[width=0.45\textwidth]{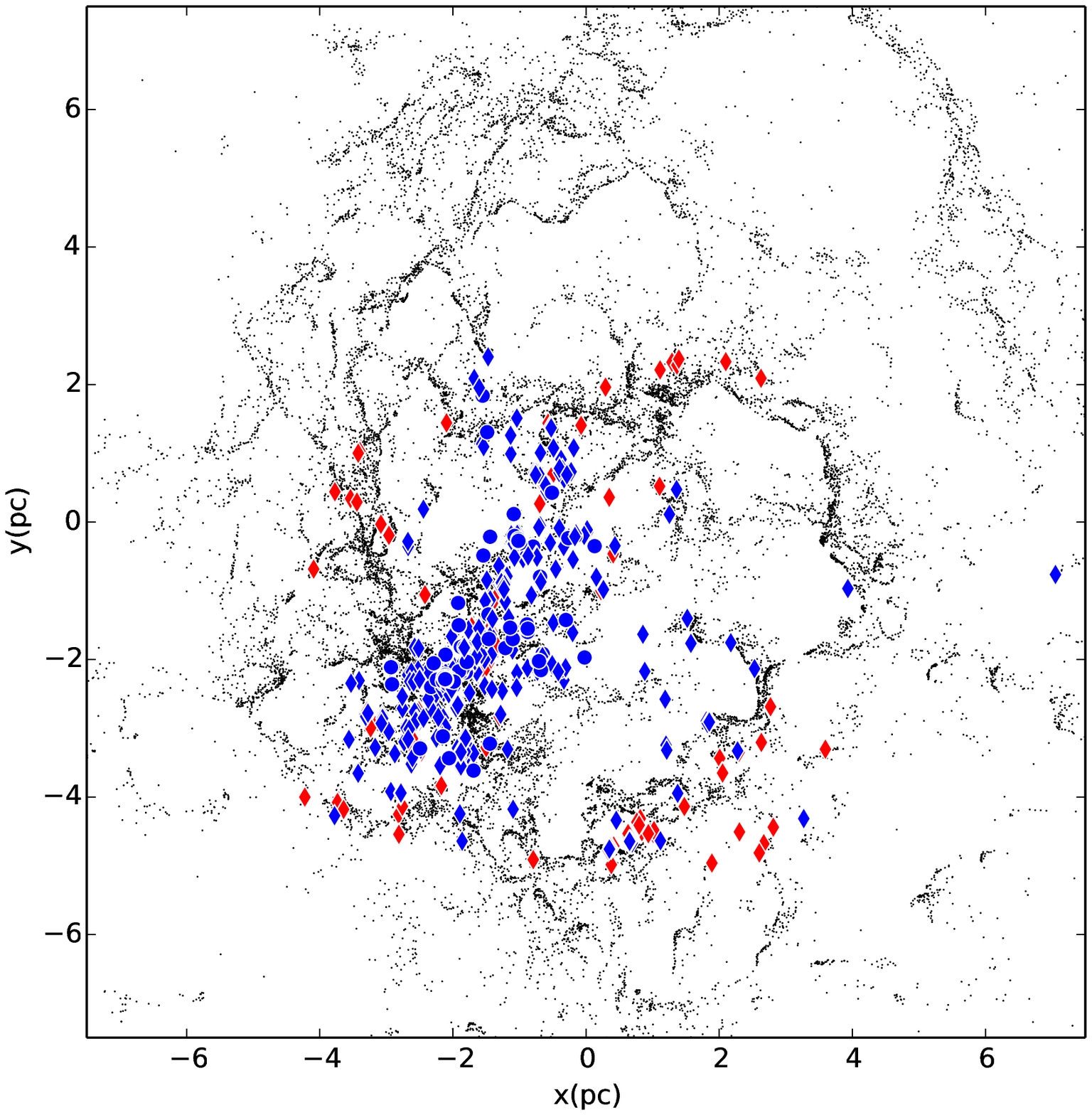}}
\vspace{0.1in}
\subfloat[Run UQ]{\includegraphics[width=0.45\textwidth]{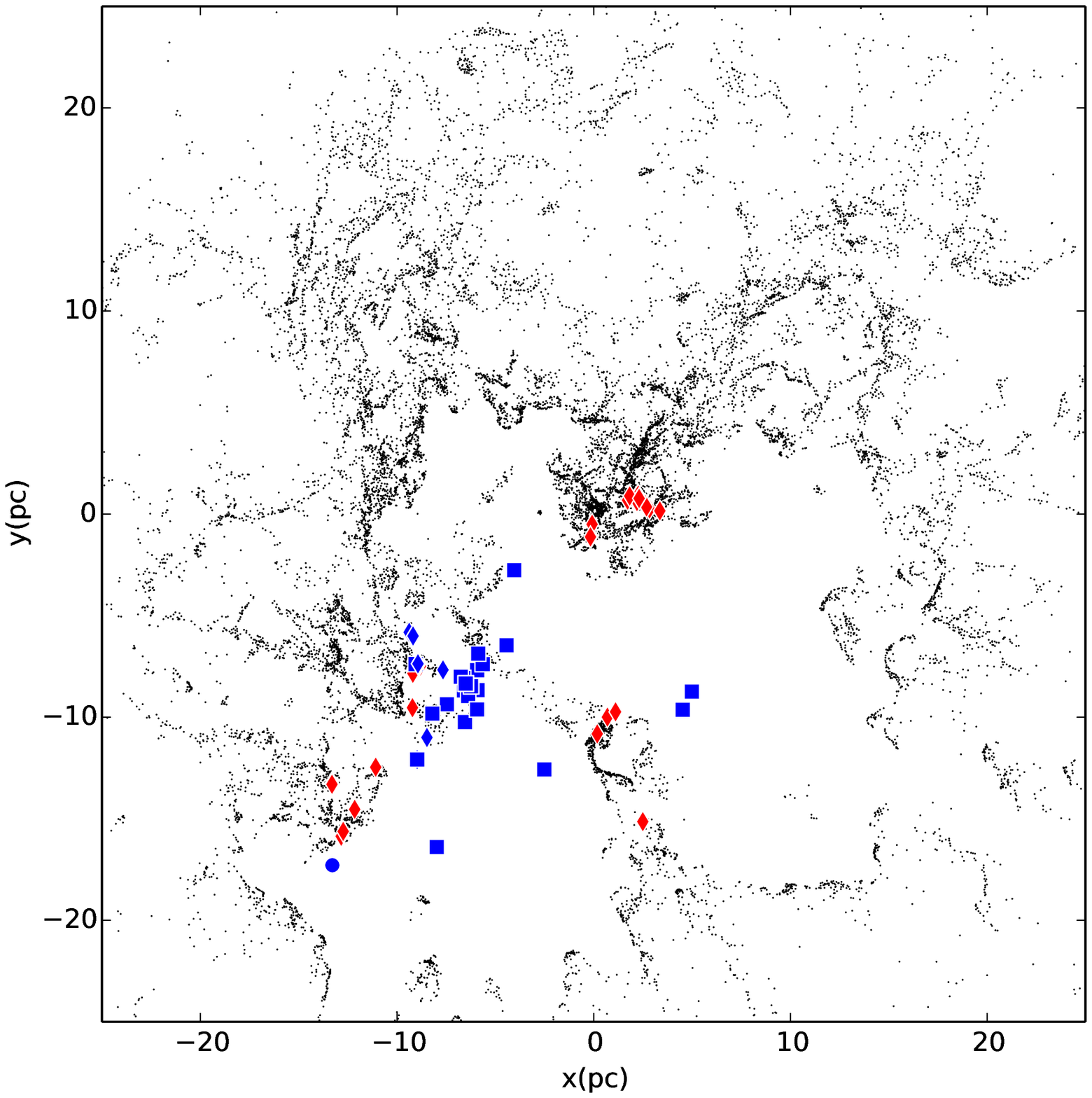}}
\hspace{0.1in}
\subfloat[Run UF]{\includegraphics[width=0.45\textwidth]{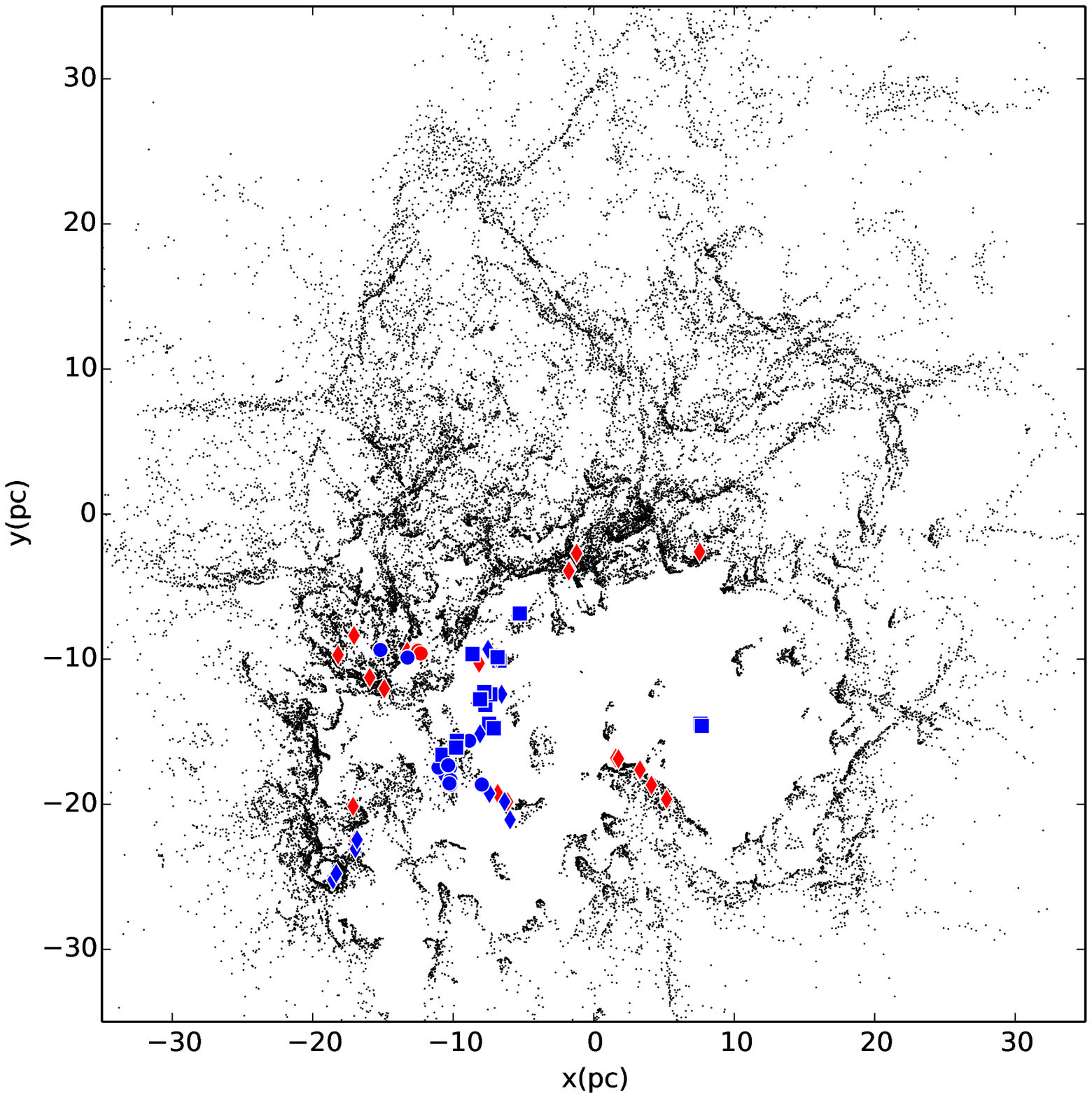}}
\caption{Relative positions of sink particles and ionisation fronts in Runs I, J, UQ and UF. Neutral gas particles on the ionisation fronts are shown as black dots. Sink particles are overlaid. Red sinks are triggered, blue sinks are spontaneously--formed, and symbol shapes denote sink ages: diamonds (age$<$1Myr); circles (1Myr$<$age$<$2Myr); squares (2Myr$<$age).}
\label{fig:ifronts}
\end{figure*}

\indent Since the edges of bubbles are not necessarily easy to define or identify, we instead make use of a better--defined marker, namely the location of the ionisation front(s). This technique is used in several of the papers assembled in Table \ref{tab:bigtable} and Figure \ref{fig:barchart}, e.g. \cite{2009ApJ...700..506S}. The use of bright--rimmed clouds as markers for triggered star formation depends essentially on the same idea, since the bright rims themselves are the glowing ionised boundary layers on the irradiated faces of the clouds.\\
\indent Considerable theoretical work has been done on the simulation of the triggered collapse of isolated clumps or cores, often modelled as Bonner--Ebert spheres \cite[e.g][]{1994A&A...289..559L,1994ApJ...420..213K,2009MNRAS.393...21G,2011ApJ...736..142B,2012MNRAS.420..562H,2012MNRAS.426..203H}. These simulations produce morphologies which strikingly resemble bright--rimmed clouds, cometary globules and pillars. \cite{2012MNRAS.426..203H} show through detailed synthetic observations of their simulations that the observational diagnostics used to analyse the evolution of BRCs are reasonably reliable, although they may underestimate the effects of shock compression. However, by their very nature, these simulations leave the origins of the globules unanswered. In common with the simulations of the collect--and--collapse process, this work models Type II and III triggering, since the initial conditions are stable in the absence of feedback.\\
\indent We now turn to the simulations of Dale et al. and examine the question of whether the location of ionisation fronts can be used as a diagnostic of triggering on larger scales, as described by \cite{2009ApJ...700..506S}. The greyscale in Figure \ref{fig:runi_gas} shows all the gas in the Run I calculation. The ionisation algorithm described in \cite{2007MNRAS.382.1759D} can be used to locate neutral particles just behind the ionisation fronts along a given ray from any radiation source. This allows us to locate the ionisation fronts in the I, J, UQ and UF calculations and test this hypothesis. In Figure \ref{fig:ifronts} we show the results of this procedure for these four calculations. Black dots denote neutral particles on the ionisation fronts and coloured symbols depict the sinks. Red sinks are triggered, blue sinks are spontaneously--formed, and symbol shapes denote sink ages: diamonds (age$<$1Myr); circles (1Myr$<$age$<$2Myr); squares (2Myr$<$age).\\
\indent In all cases, the structure and appearance of the ionisation fronts is extremely complex, partly as a result of there being multiple distributed ionising sources and partly because of the complex geometry of the clouds. There are many places in each simulation where ionisation fronts are seen face--on or at oblique angles, rather then edge--on. One might think that this would lead to severe projection effects when trying to determine the distance between a given star and the nearest ionisation front. We investigate this below. Since Runs I, UQ and UF each contain rather small numbers of stars, we group them together in a meta-analysis and compare the results with and without Run J (which contains more stars than the other three runs put together).\\
\begin{figure*}
\centering
\subfloat[All stars]{\includegraphics[width=0.45\textwidth]{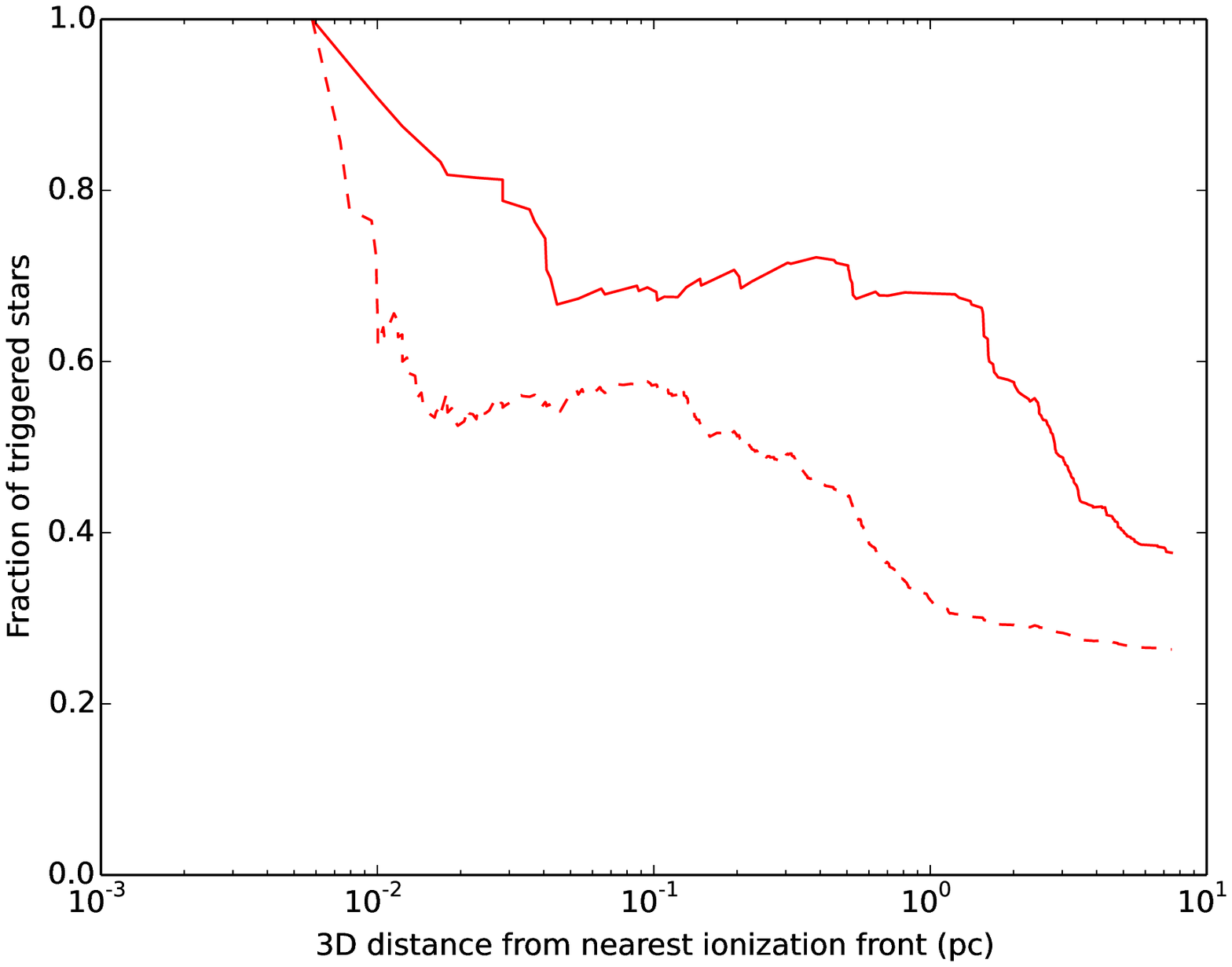}}
\hspace{0.1in}
\subfloat[Stars $<$1Myr only]{\includegraphics[width=0.45\textwidth]{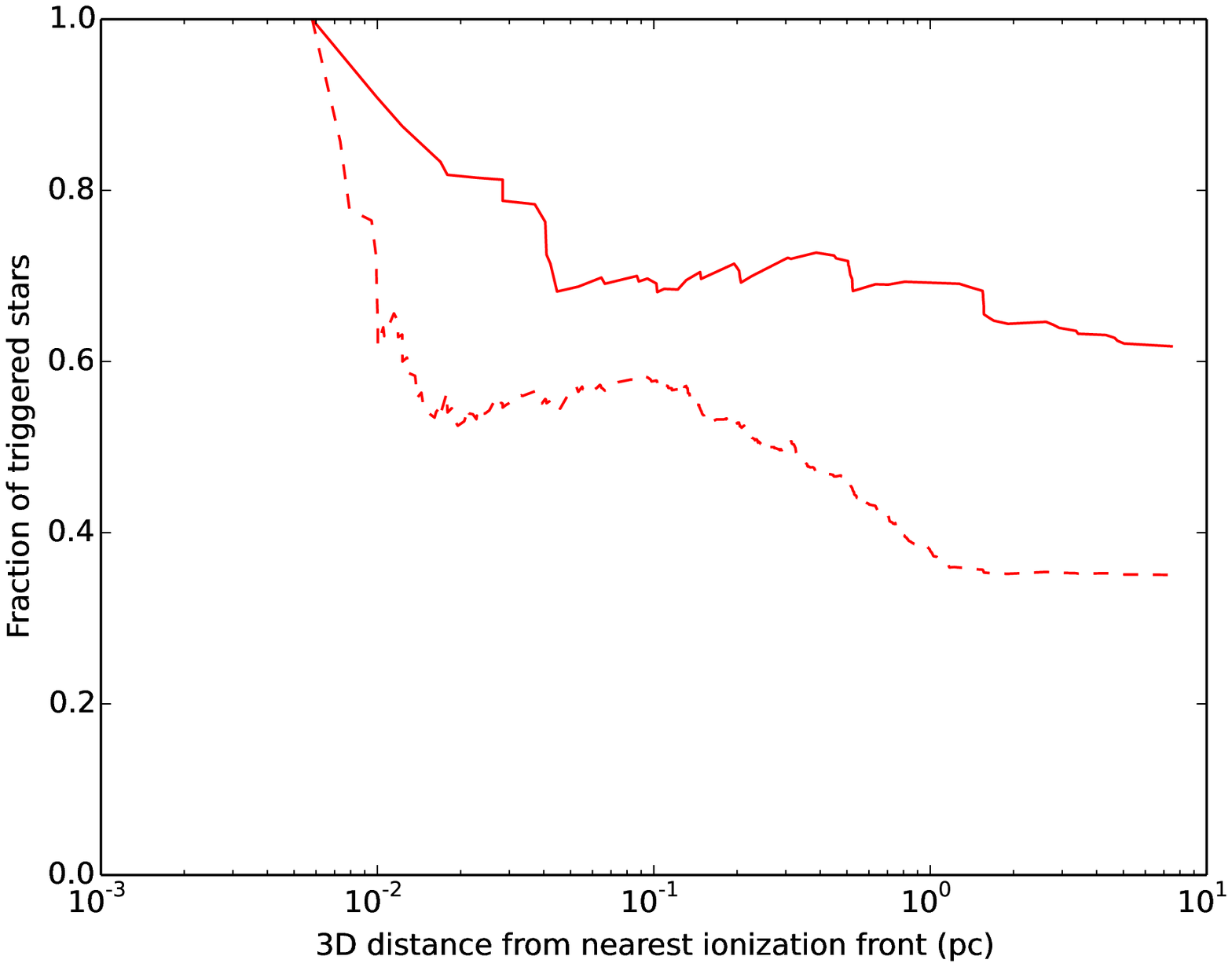}}
\caption{Fraction of triggered stars plotted against 3D separation from the nearest ionisation front. The dashed line shows the results including the very populous Run J, while the solid line excludes this calculation. All stars are included in the left panel, regardless of age, whilst only stars younger than 1Myr are included in the right panel.}
\label{fig:cuf_trig_3D}
\end{figure*}
\begin{figure*}
\centering
\subfloat[All stars]{\includegraphics[width=0.45\textwidth]{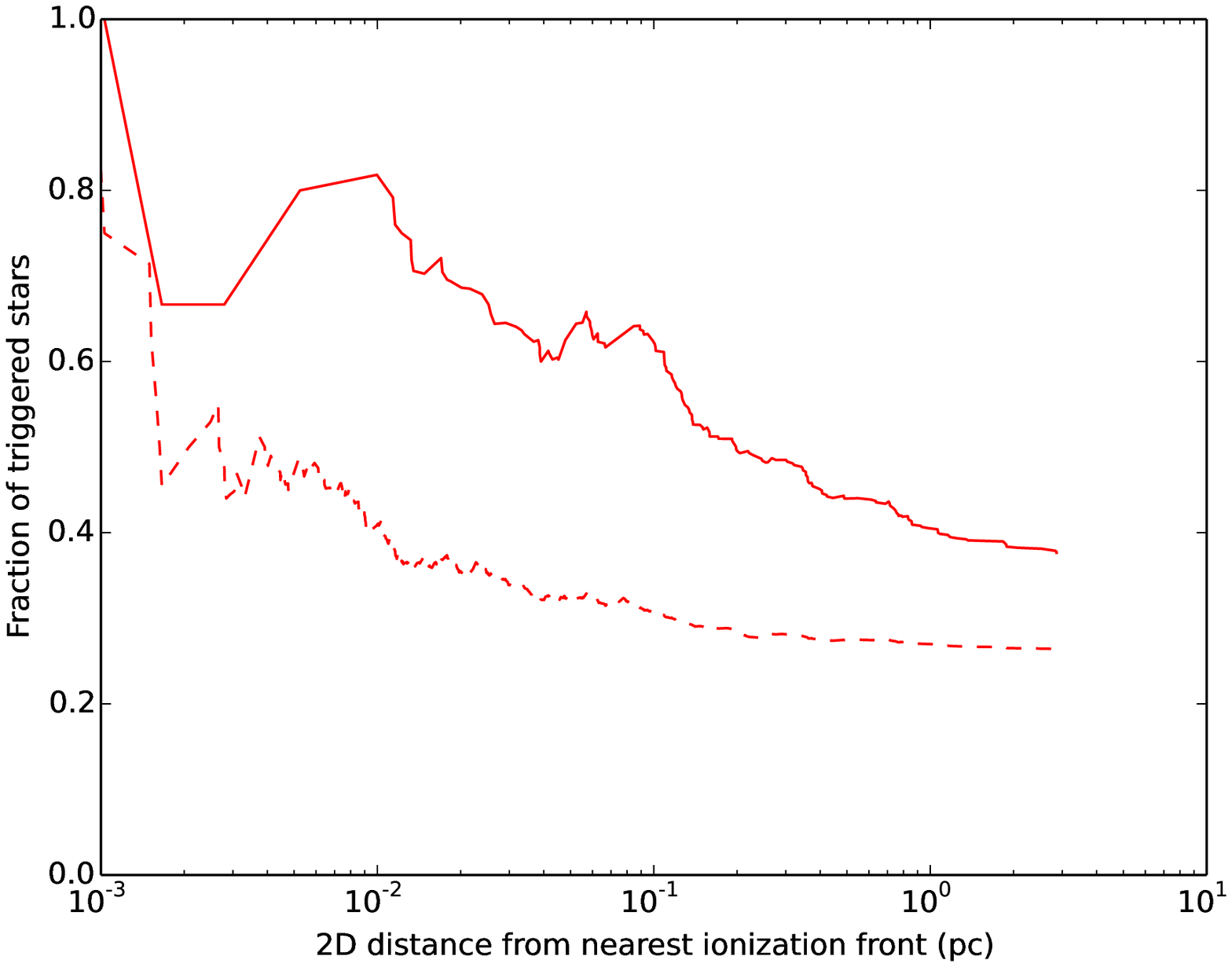}}
\hspace{0.1in}
\subfloat[Stars $<$1Myr only]{\includegraphics[width=0.45\textwidth]{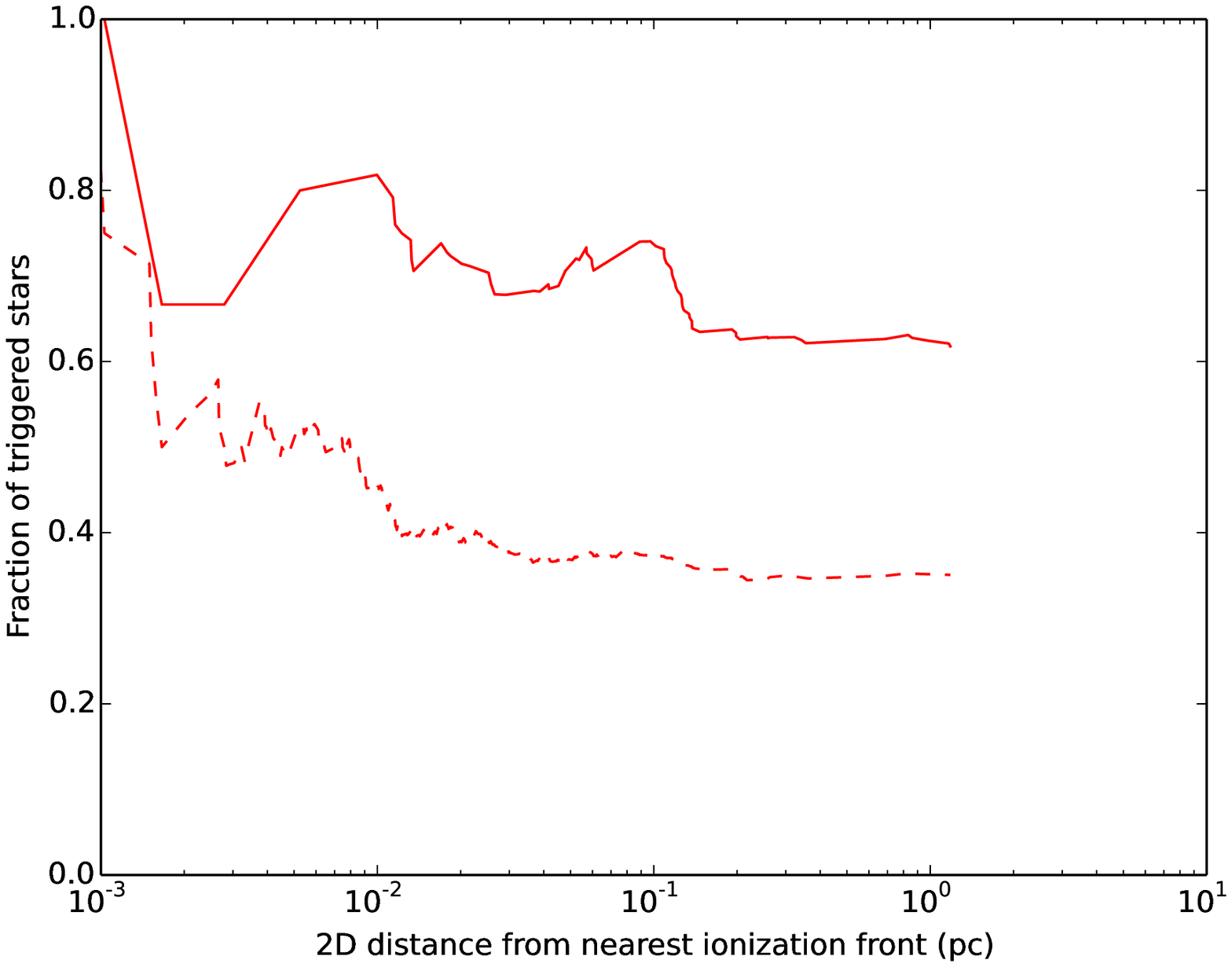}}
\caption{Fraction of triggered stars plotted against projected 2D separation from the nearest ionisation front. The dashed line shows the results including the very populous Run J, while the solid line excludes this calculation. All stars are included in the left panel, regardless of age, whilst only stars younger than 1Myr are included in the right panel.}
\label{fig:cuf_trig_2D}
\end{figure*}
\indent The justification for using proximity to an ionisation front as a criterion for deciding whether a star is triggered should be that the probability that a randomly--chosen star near an ionisation front is triggered is substantially larger than the probability that any randomly--chosen star is triggered. In the symbology of probability theory, if $P(T)$ is the probability that any random star is triggered and the probability that any random star is close to an ionisation front is $P(C)$, the requirement is that $P(T|C)\gg P(T)$.\\
\indent We can evaluate these probabilities directly from the simulations by counting how many triggered stars and how many stars in total lie within a given two-- or three-dimensional separation $s$ from the nearest ionisation front and computing\\
\begin{eqnarray}
P(T|C)=\frac{\sum_{\rm trig}(r_{\rm IF}< s)}{\sum_{\rm all}(r_{\rm IF}< s)},
\end{eqnarray}
where the numerator is the total number of \emph{triggered} stars whose perpendicular separation $r_{\rm IF}$ from the nearest ionisation front is less than $s$, and the denominator is the total number of \emph{all} stars whose perpendicular distance to the nearest ionisation front is less than $s$. $P(T|C)$ is then the fraction of stars closer than $s$ to an ionisation front that are triggered.\\
\indent Since there is no good definition of what `close' means, $s$ is allowed to range over several decades. In the left panel of Figure \ref{fig:cuf_trig_3D} we plot the fraction of triggered stars as a function of three--dimensional or projected distance $s$ from the nearest ionisation front. The solid line results form adding the results from Runs I, UQ and UF and the dashed line results from also including Run J. Note that the smallest 3D separations measured are $\approx5\times10^{-3}$pc because this is the sink particle accretion radius in these calculations.\\
\indent The total fraction of all stars in Runs I, UQ and UF which are triggered is $P(T)=0.38$, and the fraction of all stars in all four runs which are triggered is $P(T)=0.27$. The solid and dashed lines therefore tend to these numbers at large separations. These fractions represent the probability of being correct if one were to choose an object at random from the relevant simulation and simply assert that it is triggered. The purpose of this investigation is to see if using a small projected distance of a given object from an ionisation front affords any improvement over blind luck.\\
\indent The solid curve in the left panel of Figure \ref{fig:cuf_trig_3D}, representing Runs I, UQ and UF only, does rise as the ionisation front separation decreases. However, after a rapid increase between 10 and 1 pc, the improvement flattens at a probability of $P(T|C)\approx0.65$, failing to increase any further until separations of a few$\times10^{-2}$pc are reached. Using a projected proximity of between $\sim10^{-2}$ and 1pc in these runs thus improves discrimination of triggered objects over blind luck by a factor of less than two, and one would still be correct in using this criterion less than two thirds of the time. If Run J is included in the analysis, the picture is worse still. The improvement in discrimination with decreasing separation is slower and more modest, plateauing between $10^{-2}$ and $10^{-1}$ pc at $P(T|C)\approx 0.55$. This implies that using the proximity criterion in all four runs combined would still result in one being wrong about whether a given object is triggered almost half the time.\\
\indent Most triggered objects in these four calculations are less than 1Myr old. If we suppose that a putative observer had sufficiently good data to compute ages to this accuracy, they may be able to exclude older stars from their sample, in principle making triggered objects easier to spot. In the right panel of Figure \ref{fig:cuf_trig_3D}, we repeat the analysis used to generate the left panel, but exclude stars older than 1Myr. The fraction of triggered objects in stars less than this age in Runs I, UQ and UF combined is $P(T)=0.61$, and in all four runs is $P(T)=0.35$, which again give the limits to which the curves tend at large distances. In the combined I, UQ and UF data, ionisation-front proximity gives only a very modest improvement in the ability to pick out triggered objects, rising only to $P(T|C)\approx0.7$ at separations between a few$\times10^{-2}$pc and 10$^{1}$pc. Including Run J results in a greater relative improvement, up to almost $P(T|C)\approx0.6$ between $10^{-2}$ and $10^{-1}$ pc. However, we note that measuring such small separations would be difficult in practice. In addition, since we have plotted three--dimensional distances, we have made use of information that is very unlikely to be available to a real observer.\\
\indent To see what effect projection has on this problem, we simply repeat the above analysis projecting all calculations along the $z$--axis. The left and right panels of Figure \ref{fig:cuf_trig_2D} depict the result, respectively including and excluding sinks older than 1Myr. Since projected distances are being used, the separation can now be arbitrarily small. When all stars are included in the analysis, the improvement in the probability of correctly inferring triggering using proximity to an ionisation front is much slower than in the 3D case. In the combined Runs I, UQ and UF, the fraction of triggered stars increases gradually, reaching $P(T|C)\approx 0.7$ at the very small separation of $10^{-2}$pc, again roughly a factor of two improvement over blind luck. Including Run J, there is a very slim improvement from $P(T)=0.27$ to $P(T|C)\approx0.35$ at a separation of $10^{-2}$pc, implying that using even this strict criterion would still leave one being wrong about a given star being triggered almost two thirds of the time. This discouraging picture is scarcely improved by excluding the older stars, which may have moved away from the ionisation fronts, as shown in the right panel of Figure \ref{fig:cuf_trig_2D}. However, we note that these probabilities are high enough that triggering has almost certainly been genuinely identified in at least some instances in the literature surveyed.\\
\subsubsection{Pillars}
\indent 18$\%$ of papers detailed in Table \ref{tab:bigtable} and Figure \ref{fig:barchart} infer triggered star formation by the presence of stars near pillar-like objects. Pillars are distinctive and highly--photogenic structures, and have also attracted the attention of simulators. A wide variety of mechanisms have been found able to reproduce the pillar morphology, including hydrodynamic instabilities, perturbations in the density or radiation field, or the erosion of  pre--existing structures generated by turbulence \citep[e.g][]{1996ApJ...469..171G,2001MNRAS.327..788W,2010ApJ...723..971G,2012A&A...538A..31T,2012MNRAS.420..562H,2013MNRAS.435..917W}. Of these, only \cite{2013MNRAS.435..917W} actually modelled the formation of stars, finding that they were well--correlated with pillar structures in their high--fractal dimension clouds.\\
\indent We observe several pillar--like structures in the simulations of Dale at al., in particular a very prominent conical pillar from the Run I simulation, which is the eroded remains of a filamentary accretion flow. In Figure \ref{fig:runi_gas}, we show a screenshot from the end of this calculation with the gas depicted in greyscale and the sink particles shown as blue (spontaneously-formed) or red (triggered) diamonds (age$<$1Myr), circles (1Myr$<$age$<$2Myr) or squares (2Myr$<$age). There is clearly a rough age gradient, with a dense group of old stars left and below centre, with intermediate-age objects surrounding it and even younger stars surrounding these. The pillar is clearly visible in the bottom left corner. In the right panel of Figure \ref{fig:runi_gas} we show a zoom onto the pillar. Most of the stars associated with it are young, falling into the first age bracket mentioned above, but most are also not triggered. There is a dense group of stars about halfway along the pillar, several of which are triggered, but they are mixed in with their spontaneously-formed brethren.\\
\subsection{Dynamical ages, age gradients and elongated clusters}
\indent A simple first--order check to see if the triggering of a particular star is possible is to check whether the star is younger than the feedback--driving stars or the feedback--driven structure thought to be responsible. This technique is used in $\approx37\%$ of the papers cited in Table \ref{tab:bigtable} and Figure \ref{fig:barchart}. Determining the age of an O--star or an expanding HII region or wind bubble are subject to uncertainties and are unlikely to be accurate to better than 1Myr. In the simulations of Dale et al., the triggered and spontaneously--formed stars are geometrically mixed. The HII regions in these simulations are 1.5--3 Myr old and the massive stars driving them are therefore substantially older than that. We tested in the previous subsection to see whether restricting the analysis to only stars $<$1Myr old, and therefore assuredly younger than the O--stars or the HII regions helped in the detection of triggered stars. We found that it did, but not substantially.\\
\indent \cite{2013MNRAS.431.1062D} examined the usefulness of spatial age gradients in distinguishing triggered stars from their spontaneously--formed colleagues and found that they were of limited help. They employed two possible measures of the ages of the sink particles -- the time since they first formed and the time since they ceased accreting and acquired their final masses. They found that the former definition resulted in no identifiable age gradients. The latter definition did produce age gradients in some simulations but in both classes of object. As the ionisation front(s) washed over a given region, they were able to cut off the supply of gas to triggered and spontaneously--formed stars alike, leading to age gradients in both species. The simulations exhibit no evidence of geometric elongation of clusters towards feedback sources, but even if there were, it would not serve to distinguish the triggered and spontaneous stars. The shocks driven by the ionisation fronts are also able to redistribute spontaneously--forming objects amongst the triggered stars, further blurring the distinction between them. \cite{2013MNRAS.435..917W} observed clear age gradients in some of their clouds, most strongly in clouds with larger fractal dimensions. However, since their control simulations do not form any stars over the dynamical ages of the HII regions, mixing with spontaneously--formed stars is not an issue in their simulations.\\
\section{Discussion}
\subsection{Insights into triggering from simulations of turbulent clouds}
Star formation triggered by other stars within individual molecular clouds is a tantalising possibility and has attracted much theoretical, numerical and observational interest. It is relatively easy to define and detect triggering in theoretical work, most easily by starting from stable initial conditions, or by comparison with control simulations without feedback. These luxuries are in general not available to observers. Instead, a range of signposts have been devised and, singly or in combination, used to infer the action of triggering, in the sense of creating larger numbers of stars, in many star--forming regions. Detailed numerical simulations are, however, now able to reproduce all of these signposts and unfortunately, it appears that they are by no means foolproof.\\
\indent We have here extended the analysis presented in the recent series of papers by Dale et al of the influence of ionising feedback on turbulent GMCs. The use of a Lagrangian hydrodynamics scheme and control simulations without feedback allowed Dale et al to show unequivocally that triggered star formation was present in their simulations. However, the triggered and spontaneously--formed stars in these calculations were spatially mixed, making the two populations hard to distinguish. We examined in detail several correlations which have been used by observers to sort triggered from spontaneous stars on a local star-by-star basis to see how they perform in the context of the simulations.\\
\indent Most striking is the failure of the relative locations of stars and shells, pillars, or ionisation fronts to provide substantial assistance in discriminating triggered from non-triggered objects. This is a particularly important result, since these techniques are those most commonly used by observers (see Table \ref{tab:bigtable} and Figure \ref{fig:barchart}). Runs I, UQ and UF have relative simple geometries in which the gas structure is characterised by one or a few well-defined bubbles. Even if stellar positions and ionisation front locations were available in three dimensions to accuracies of $\sim$10$^{-2}$pc, using this criterion gives an improvement of only a factor of less than two over blind luck. Eliminating stars older than 1Myr, on the grounds that virtually all triggered objects in these calculations are younger than this age while many spontaneously-formed stars are older, helps somewhat but only allows triggered stars to be reliably identified about two thirds of the time. If projection effects are taken into account, the success rates become much worse.\\
\subsection{Why triggering is so hard to observe}
\begin{figure*}
\centering
\subfloat[Run I]{\includegraphics[width=0.45\textwidth]{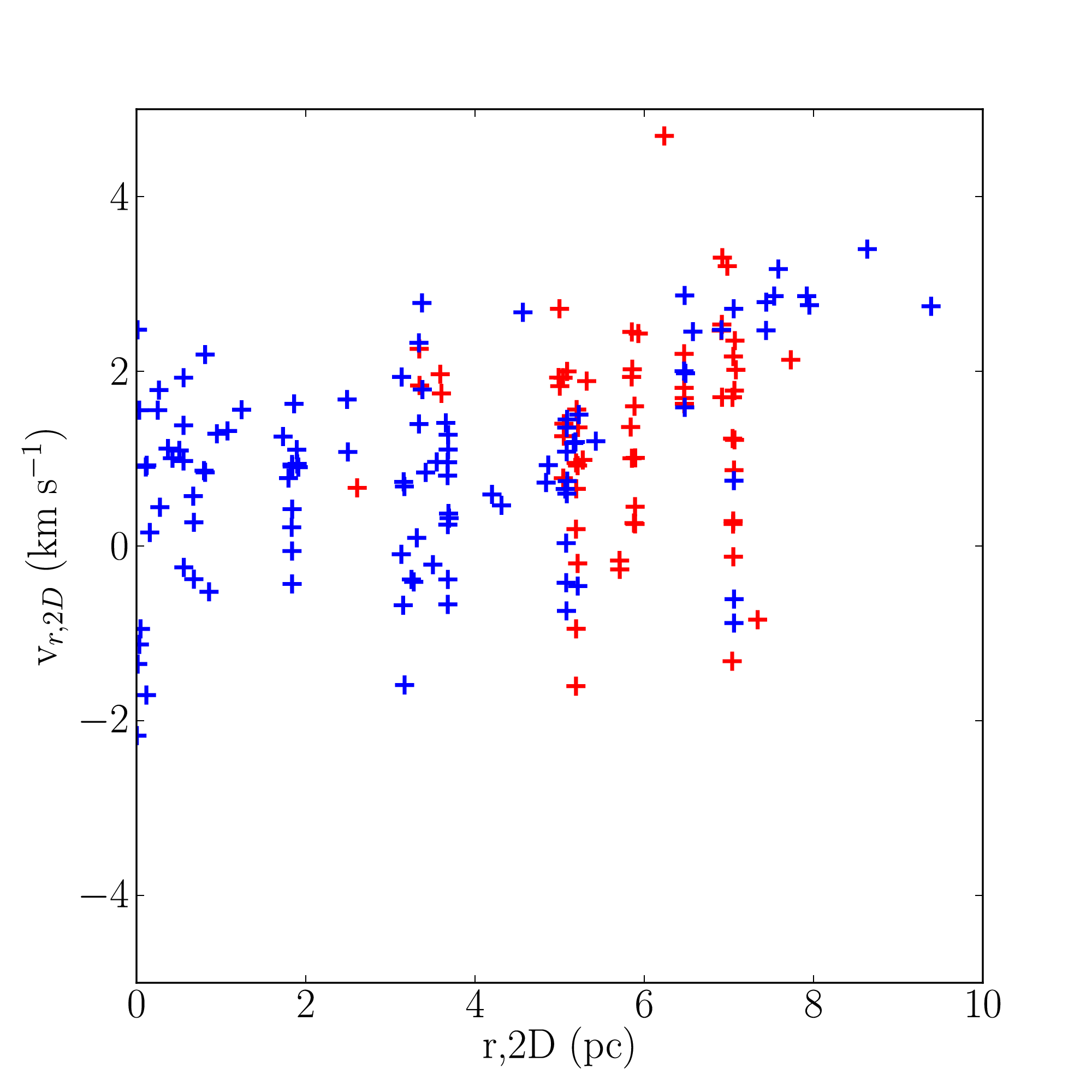}}
\hspace{0.1in}
\subfloat[Run UQ]{\includegraphics[width=0.45\textwidth]{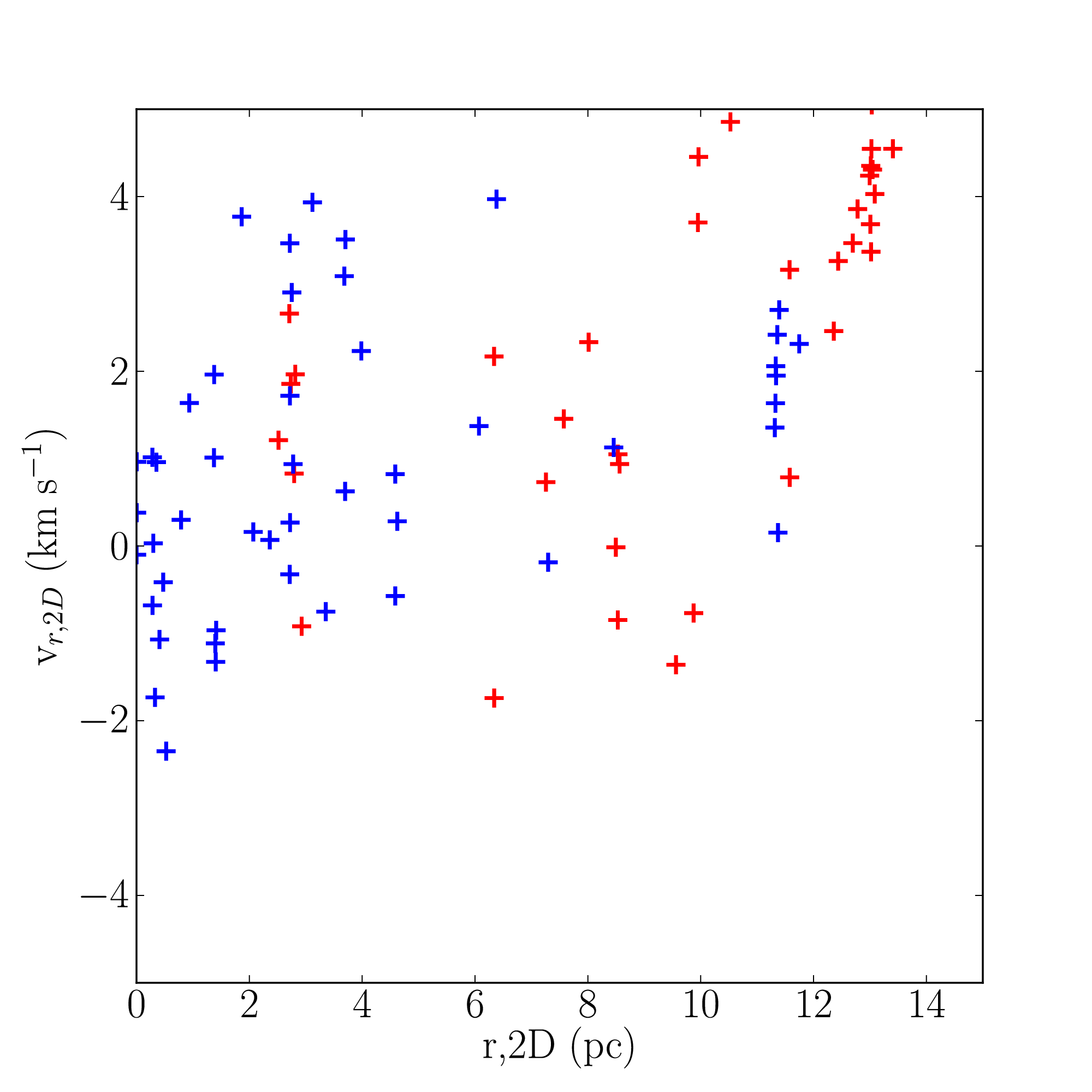}}
\caption{Projected velocity plotted against projected distance, both with respect to the most massive ionising star, for spontaneously--formed objects (blue) and triggered objects (red) in Run I (left panel) and Run UQ (right panel).}
\label{fig:r2dvr2d}
\end{figure*}

\indent The reason for these outwardly surprising results was adumbrated in \cite{2013MNRAS.431.1062D}. As well as triggering and aborting star formation, feedback \emph{redistributes} star formation. Expanding ionisation fronts collect material, some of which was going to form stars anyway, or is already in the process of forming stars, and some of which was not, and moves it to a different location. Star formation at the new location is thus likely to involve stars that were going to form anyway as well as triggered stars, and both types of object are therefore likely to be found near feedback-driven structures such as pillars, bubble perimeters or ionisation fronts. Additionally, the expansion of HII regions can be locally arrested by running into dense obstacles in which star formation was already imminent or underway, again resulting in spatial correlation of spontaneously-formed stars with signposts of feedback.\\
\indent A detailed knowledge of the dynamical states of the stars might be thought to alleviate this problem. A natural consequence of triggered star formation in a given region should be that the triggered objects are moving approximately radially away from the older massive stars whose feedback is doing the triggering. Triggered objects projected near the massive stars in the sky should then have large line-of-sight velocities, and those at large projected distance should have large proper motions directed away from the O-stars. Both of these are in principle measurable from observations of the stars alone.\\
\indent However, the redistributive effect operates in velocity space as well as real space. In Figure \ref{fig:r2dvr2d}, we show the magnitude of the proper motion of the sinks from the Runs I and UQ calculations with respect, for each sink, to the most massive (and in both cases oldest) ionising source as a function of projected distance from that source. Blue symbols denote spontaneously-formed sinks and red symbols triggered sinks. In Run UQ, there is a clear population of $\sim$ ten triggered objects at both large radii and large proper motion with respect to the most massive star.\\
\indent There are, though, spontaneously-formed stars in a similar region of the diagram at slightly smaller radii and proper motions. In addition, there are other triggered objects mixed in with spontaneous objects throughout the rest of the plot, including triggered stars which are moving towards the ionising source. The corresponding plot from Run I shows no clear demarcation between triggered and spontaneous objects. We also compared line-of-sight velocities against projected distance from the ionising stars, but there are no visible correlations in these plots. There is some tendency for the triggered objects to have larger positive radial proper motions correlated with larger projected distance from the most massive stars, but not a strong one. Some spontaneously-formed stars acquire not only similar positions but also similar velocities to the triggered stars.\\
\section{Conclusions}
\indent Triggered star formation may occur on many different scales and be due to many different triggering agents. On the largest scales, star formation can be triggered on the scales of whole GMCs by external agents such as galactic collisions. Then, one has synchronised star formation over very large distances, and the collision itself is an obvious candidate triggering process. In such clear--cut cases, inferring triggering is then relatively straightforward and secure. On scales smaller than a whole GMC, the inference of internal triggering becomes more difficult. This is largely because, by definition, star formation must already be well underway before this type of triggering can occur. Distinguishing triggered from spontaneously-formed stars is not easy and, given that the region in question would form more stars even in the absence of feedback, it is very hard to say how the final result of the star formation process is likely to be affected.\\
\indent Triggered star formation can have several different meanings, such as increasing the star formation efficiency, increasing the star formation rate or increasing the total numbers of stars formed. Which definition is intended is often not clear in a given item in the literature, but triggering is usually inferred by correlating particular stars with some feedback--driven structure. The ages and distribution of stars relative to the feedback source or feedback--driven structure are also often used to show that triggering is plausible. \\
\indent By contrast, demonstrating triggered star formation in numerical simulations is relatively easy. This allows the fidelity of the observational techniques to be tested. We found that none of the correlations with shells, ionisation fronts or pillar structures were of substantial help in winnowing out the triggered objects. Neither were relative ages or geometrical distribution of stars. The source of this failure is the redistribution of spontaneous star formation to the same locations in position and velocity space as triggered stars.\\
\indent We therefore feel that, despite the wealth of data available and the sophisticated analyses applied to it, statements made about triggered star formation should be interpreted with great care, especially when they refer to small numbers of individual stars. Of course, systems where many putative triggering indicators can be satisfied simultaneously are more likely to be genuine sites of triggering.\\
\indent Simulations and theory suggest that the overall effects of feedback on the star formation process are most likely negative in terms of the rates at which gas is converted to stars or the final star formation efficiency of a given system. However, such a statement only makes sense against the backdrop of a counterpart `control' system in which feedback is absent. Since no such systems exist in reality, the terms `triggered star formation' and `aborted star formation' are very hard to define outside controlled artificial environments.\\
\indent Stellar feedback is an integral part of the star formation process, in the same way as gravitational collapse or accretion, and the effects of feedback can certainly be teased out. There are clearly star-forming regions, such as Orion or W3/4/5, whose structure and dynamics (both stellar and gaseous) are due largely to O--star feedback, and conversely there are regions such as Taurus that lack massive stars for which this is obviously not the case. Star formation in systems belonging to the former category could perhaps be described as `feedback-dominated', `feedback-governed' or at least `feedback-influenced'. Some of the effects of feedback in such systems could likely be inferred, but speculating on how they would look and evolve on large scales \emph{in the absence of feedback}, which is essential for the terms `triggered' or `aborted' to be meaningful, is extremely difficult in an observational context.\\
\section{Acknowledgements}
This research was supported by the DFG cluster of excellence `Origin and Structure of the Universe' (JED). The authors thank the referee, J. Palou\v{s}, for a very swift report that substantially improved the clarity of the paper.

\bibliography{myrefs}

\begin{thebibliography}{}

\bibitem[\protect\citeauthoryear{{Bate}, {Bonnell} \& {Price}}{{Bate}
  et~al.}{1995}]{1995MNRAS.277..362B}
{Bate} M.~R.,  {Bonnell} I.~A.,    {Price} N.~M.,  1995, \mnras, 277, 362

\bibitem[\protect\citeauthoryear{{Beltr{\'a}n}, {Massi}, {L{\'o}pez}, {Girart}
  \& {Estalella}}{{Beltr{\'a}n} et~al.}{2009}]{2009A&A...504...97B}
{Beltr{\'a}n} M.~T.,  {Massi} F.,  {L{\'o}pez} R.,  {Girart} J.~M.,
  {Estalella} R.,  2009, \aap, 504, 97

\bibitem[\protect\citeauthoryear{{Benz}}{{Benz}}{1990}]{1990nmns.work..269B}
{Benz} W.,  1990, in {Buchler} J.~R.,  ed., Numerical Modelling of Nonlinear
  Stellar Pulsations Problems and Prospects {Smooth Particle Hydrodynamics - a
  Review}.
pp 269--+

\bibitem[\protect\citeauthoryear{{Bieging}, {Peters}, {Vila Vilaro},
  {Schlottman} \& {Kulesa}}{{Bieging} et~al.}{2009}]{2009AJ....138..975B}
{Bieging} J.~H.,  {Peters} W.~L.,  {Vila Vilaro} B.,  {Schlottman} K.,
  {Kulesa} C.,  2009, \aj, 138, 975

\bibitem[\protect\citeauthoryear{{Bik}, {Puga}, {Waters}, {Horrobin},
  {Henning}, {Vasyunina}, {Beuther}, {Linz} \& {Kaper}}{{Bik}
  et~al.}{2010}]{2010ApJ...713..883B}
{Bik} A.,  {Puga} E.,  {Waters} L.~B.~F.~M.,  {Horrobin} M.,  {Henning} T.,
  {Vasyunina} T.,  {Beuther} H.,  {Linz} H.,    {Kaper} L.,  2010, \apj, 713,
  883

\bibitem[\protect\citeauthoryear{{Billot}, {Noriega-Crespo}, {Carey}, {Guieu},
  {Shenoy}, {Paladini} \& {Latter}}{{Billot}
  et~al.}{2010}]{2010ApJ...712..797B}
{Billot} N.,  {Noriega-Crespo} A.,  {Carey} S.,  {Guieu} S.,  {Shenoy} S.,
  {Paladini} R.,    {Latter} W.,  2010, \apj, 712, 797

\bibitem[\protect\citeauthoryear{{Bisbas}, {W{\"u}nsch}, {Whitworth}, {Hubber}
  \& {Walch}}{{Bisbas} et~al.}{2011}]{2011ApJ...736..142B}
{Bisbas} T.~G.,  {W{\"u}nsch} R.,  {Whitworth} A.~P.,  {Hubber} D.~A.,
  {Walch} S.,  2011, \apj, 736, 142

\bibitem[\protect\citeauthoryear{{Cambr{\'e}sy}, {Marton}, {Feher}, {T{\'o}th}
  \& {Schneider}}{{Cambr{\'e}sy} et~al.}{2013}]{2013A&A...557A..29C}
{Cambr{\'e}sy} L.,  {Marton} G.,  {Feher} O.,  {T{\'o}th} L.~V.,    {Schneider}
  N.,  2013, \aap, 557, A29

\bibitem[\protect\citeauthoryear{{Chauhan}, {Pandey}, {Ogura}, {Jose}, {Ojha},
  {Samal} \& {Mito}}{{Chauhan} et~al.}{2011}]{2011MNRAS.415.1202C}
{Chauhan} N.,  {Pandey} A.~K.,  {Ogura} K.,  {Jose} J.,  {Ojha} D.~K.,  {Samal}
  M.~R.,    {Mito} H.,  2011, \mnras, 415, 1202

\bibitem[\protect\citeauthoryear{{Chauhan}, {Pandey}, {Ogura}, {Ojha}, {Bhatt},
  {Ghosh} \& {Rawat}}{{Chauhan} et~al.}{2009}]{2009MNRAS.396..964C}
{Chauhan} N.,  {Pandey} A.~K.,  {Ogura} K.,  {Ojha} D.~K.,  {Bhatt} B.~C.,
  {Ghosh} S.~K.,    {Rawat} P.~S.,  2009, \mnras, 396, 964

\bibitem[\protect\citeauthoryear{{Choudhury}, {Mookerjea} \&
  {Bhatt}}{{Choudhury} et~al.}{2010}]{2010ApJ...717.1067C}
{Choudhury} R.,  {Mookerjea} B.,    {Bhatt} H.~C.,  2010, \apj, 717, 1067

\bibitem[\protect\citeauthoryear{{Churchwell}, {Povich}, {Allen}, {Taylor},
  {Meade}, {Babler}, {Indebetouw}, {Watson}, {Whitney}, {Wolfire}, {Bania},
  {Benjamin}, {Clemens}, {Cohen}, {Cyganowski}, {Jackson}, {Kobulnicky} \&
  {Mathis}}{{Churchwell} et~al.}{2006}]{2006ApJ...649..759C}
{Churchwell} E.,  {Povich} M.~S.,  {Allen} D.,  {Taylor} M.~G.,  {Meade} M.~R.,
   {Babler} B.~L.,  {Indebetouw} R.,  {Watson} C.,  {Whitney} B.~A.,  {Wolfire}
  M.~G.,  {Bania} T.~M.,  {Benjamin} R.~A.,  {Clemens} D.~P.,  {Cohen} M.,
  {Cyganowski} C.~J.,  {Jackson} J.~M.,  {Kobulnicky} H.~A.,    {Mathis} J.~S.,
   2006, \apj, 649, 759

\bibitem[\protect\citeauthoryear{{Churchwell}, {Watson}, {Povich}, {Taylor},
  {Babler}, {Meade}, {Benjamin}, {Indebetouw} \& {Whitney}}{{Churchwell}
  et~al.}{2007}]{2007ApJ...670..428C}
{Churchwell} E.,  {Watson} D.~F.,  {Povich} M.~S.,  {Taylor} M.~G.,  {Babler}
  B.~L.,  {Meade} M.~R.,  {Benjamin} R.~A.,  {Indebetouw} R.,    {Whitney}
  B.~A.,  2007, \apj, 670, 428

\bibitem[\protect\citeauthoryear{{Cichowolski}, {Pineault}, {Gamen}, {Arnal},
  {Suad} \& {Ortega}}{{Cichowolski} et~al.}{2014}]{2014MNRAS.438.1089C}
{Cichowolski} S.,  {Pineault} S.,  {Gamen} R.,  {Arnal} E.~M.,  {Suad} L.~A.,
   {Ortega} M.~E.,  2014, \mnras, 438, 1089

\bibitem[\protect\citeauthoryear{{Clark} \& {Porter}}{{Clark} \&
  {Porter}}{2004}]{2004A&A...427..839C}
{Clark} J.~S.,  {Porter} J.~M.,  2004, \aap, 427, 839

\bibitem[\protect\citeauthoryear{{Dale} \& {Bonnell}}{{Dale} \&
  {Bonnell}}{2011}]{2011MNRAS.414..321D}
{Dale} J.~E.,  {Bonnell} I.,  2011, \mnras, 414, 321

\bibitem[\protect\citeauthoryear{{Dale}, {Bonnell}, {Clarke} \& {Bate}}{{Dale}
  et~al.}{2005}]{2005MNRAS.358..291D}
{Dale} J.~E.,  {Bonnell} I.~A.,  {Clarke} C.~J.,    {Bate} M.~R.,  2005,
  \mnras, 358, 291

\bibitem[\protect\citeauthoryear{{Dale}, {Bonnell} \& {Whitworth}}{{Dale}
  et~al.}{2007}]{2007MNRAS.375.1291D}
{Dale} J.~E.,  {Bonnell} I.~A.,    {Whitworth} A.~P.,  2007, \mnras, 375, 1291

\bibitem[\protect\citeauthoryear{{Dale}, {Clark} \& {Bonnell}}{{Dale}
  et~al.}{2007}]{2007MNRAS.377..535D}
{Dale} J.~E.,  {Clark} P.~C.,    {Bonnell} I.~A.,  2007, \mnras, 377, 535

\bibitem[\protect\citeauthoryear{{Dale}, {Ercolano} \& {Bonnell}}{{Dale}
  et~al.}{2012a}]{2012MNRAS.427.2852D}
{Dale} J.~E.,  {Ercolano} B.,    {Bonnell} I.~A.,  2012a, \mnras, 427, 2852

\bibitem[\protect\citeauthoryear{{Dale}, {Ercolano} \& {Bonnell}}{{Dale}
  et~al.}{2012b}]{2012MNRAS.424..377D}
{Dale} J.~E.,  {Ercolano} B.,    {Bonnell} I.~A.,  2012b, \mnras, 424, 377

\bibitem[\protect\citeauthoryear{{Dale}, {Ercolano} \& {Bonnell}}{{Dale}
  et~al.}{2013a}]{2013MNRAS.431.1062D}
{Dale} J.~E.,  {Ercolano} B.,    {Bonnell} I.~A.,  2013a, \mnras, 431, 1062

\bibitem[\protect\citeauthoryear{{Dale}, {Ercolano} \& {Bonnell}}{{Dale}
  et~al.}{2013b}]{2013MNRAS.430..234D}
{Dale} J.~E.,  {Ercolano} B.,    {Bonnell} I.~A.,  2013b, \mnras, 430, 234

\bibitem[\protect\citeauthoryear{{Dale}, {Ercolano} \& {Clarke}}{{Dale}
  et~al.}{2007}]{2007MNRAS.382.1759D}
{Dale} J.~E.,  {Ercolano} B.,    {Clarke} C.~J.,  2007, \mnras, 382, 1759

\bibitem[\protect\citeauthoryear{{Dale}, {Ngoumou}, {Ercolano} \&
  {Bonnell}}{{Dale} et~al.}{2014}]{2014MNRAS.442..694D}
{Dale} J.~E.,  {Ngoumou} J.,  {Ercolano} B.,    {Bonnell} I.~A.,  2014, \mnras,
  442, 694

\bibitem[\protect\citeauthoryear{{Dale}, {W{\"u}nsch}, {Smith}, {Whitworth} \&
  {Palou{\v s}}}{{Dale} et~al.}{2011}]{2011MNRAS.411.2230D}
{Dale} J.~E.,  {W{\"u}nsch} R.,  {Smith} R.~J.,  {Whitworth} A.,    {Palou{\v
  s}} J.,  2011, \mnras, 411, 2230

\bibitem[\protect\citeauthoryear{{Dawson}, {McClure-Griffiths}, {Dickey} \&
  {Fukui}}{{Dawson} et~al.}{2011}]{2011ApJ...741...85D}
{Dawson} J.~R.,  {McClure-Griffiths} N.~M.,  {Dickey} J.~M.,    {Fukui} Y.,
  2011, \apj, 741, 85

\bibitem[\protect\citeauthoryear{{Deharveng}, {Lefloch}, {Kurtz}, {Nadeau},
  {Pomar{\`e}s}, {Caplan} \& {Zavagno}}{{Deharveng}
  et~al.}{2008}]{2008A&A...482..585D}
{Deharveng} L.,  {Lefloch} B.,  {Kurtz} S.,  {Nadeau} D.,  {Pomar{\`e}s} M.,
  {Caplan} J.,    {Zavagno} A.,  2008, \aap, 482, 585

\bibitem[\protect\citeauthoryear{{Deharveng}, {Lefloch}, {Massi}, {Brand},
  {Kurtz}, {Zavagno} \& {Caplan}}{{Deharveng}
  et~al.}{2006}]{2006A&A...458..191D}
{Deharveng} L.,  {Lefloch} B.,  {Massi} F.,  {Brand} J.,  {Kurtz} S.,
  {Zavagno} A.,    {Caplan} J.,  2006, \aap, 458, 191

\bibitem[\protect\citeauthoryear{{Deharveng}, {Lefloch}, {Zavagno}, {Caplan},
  {Whitworth}, {Nadeau} \& {Mart{\'{\i}}n}}{{Deharveng}
  et~al.}{2003}]{2003A&A...408L..25D}
{Deharveng} L.,  {Lefloch} B.,  {Zavagno} A.,  {Caplan} J.,  {Whitworth} A.~P.,
   {Nadeau} D.,    {Mart{\'{\i}}n} S.,  2003, \aap, 408, L25

\bibitem[\protect\citeauthoryear{{Deharveng}, {Schuller}, {Anderson},
  {Zavagno}, {Wyrowski}, {Menten}, {Bronfman}, {Testi}, {Walmsley} \&
  {Wienen}}{{Deharveng} et~al.}{2010}]{2010A&A...523A...6D}
{Deharveng} L.,  {Schuller} F.,  {Anderson} L.~D.,  {Zavagno} A.,  {Wyrowski}
  F.,  {Menten} K.~M.,  {Bronfman} L.,  {Testi} L.,  {Walmsley} C.~M.,
  {Wienen} M.,  2010, \aap, 523, A6

\bibitem[\protect\citeauthoryear{{Deharveng}, {Zavagno} \&
  {Caplan}}{{Deharveng} et~al.}{2005}]{2005A&A...433..565D}
{Deharveng} L.,  {Zavagno} A.,    {Caplan} J.,  2005, \aap, 433, 565

\bibitem[\protect\citeauthoryear{{Deharveng}, {Zavagno}, {Salas}, {Porras},
  {Caplan} \& {Cruz-Gonz{\'a}lez}}{{Deharveng}
  et~al.}{2003}]{2003A&A...399.1135D}
{Deharveng} L.,  {Zavagno} A.,  {Salas} L.,  {Porras} A.,  {Caplan} J.,
  {Cruz-Gonz{\'a}lez} I.,  2003, \aap, 399, 1135

\bibitem[\protect\citeauthoryear{{Deharveng}, {Zavagno}, {Schuller}, {Caplan},
  {Pomar{\`e}s} \& {De Breuck}}{{Deharveng} et~al.}{2009}]{2009A&A...496..177D}
{Deharveng} L.,  {Zavagno} A.,  {Schuller} F.,  {Caplan} J.,  {Pomar{\`e}s} M.,
     {De Breuck} C.,  2009, \aap, 496, 177

\bibitem[\protect\citeauthoryear{{Dewangan} \& {Ojha}}{{Dewangan} \&
  {Ojha}}{2013}]{2013MNRAS.429.1386D}
{Dewangan} L.~K.,  {Ojha} D.~K.,  2013, \mnras, 429, 1386

\bibitem[\protect\citeauthoryear{{Dewangan}, {Ojha}, {Anandarao}, {Ghosh} \&
  {Chakraborti}}{{Dewangan} et~al.}{2012}]{2012ApJ...756..151D}
{Dewangan} L.~K.,  {Ojha} D.~K.,  {Anandarao} B.~G.,  {Ghosh} S.~K.,
  {Chakraborti} S.,  2012, \apj, 756, 151

\bibitem[\protect\citeauthoryear{{Egorov}, {Lozinskaya}, {Moiseev} \&
  {Smirnov-Pinchukov}}{{Egorov} et~al.}{2014}]{2014MNRAS.444..376E}
{Egorov} O.~V.,  {Lozinskaya} T.~A.,  {Moiseev} A.~V.,    {Smirnov-Pinchukov}
  G.~V.,  2014, \mnras, 444, 376

\bibitem[\protect\citeauthoryear{{Elmegreen}}{{Elmegreen}}{1998}]{1998ASPC..148..150E}
{Elmegreen} B.~G.,  1998, in {Woodward} C.~E.,  {Shull} J.~M.,   {Thronson} Jr.
  H.~A.,  eds, Origins Vol.~148 of Astronomical Society of the Pacific
  Conference Series, {Observations and Theory of Dynamical Triggers for Star
  Formation}.
p.~150

\bibitem[\protect\citeauthoryear{{Elmegreen} \& {Lada}}{{Elmegreen} \&
  {Lada}}{1977}]{1977ApJ...214..725E}
{Elmegreen} B.~G.,  {Lada} C.~J.,  1977, \apj, 214, 725

\bibitem[\protect\citeauthoryear{{Evans} II, {Dunham}, {J{\o}rgensen}, {Enoch}
  \& {Mer{\'{\i}}n}}{{Evans} et~al.}{2009}]{2009ApJS..181..321E}
{Evans} II N.~J.,  {Dunham} M.~M.,  {J{\o}rgensen} J.~K.,  {Enoch} M.~L.,
  {Mer{\'{\i}}n} B.,  2009, \apjs, 181, 321

\bibitem[\protect\citeauthoryear{{Fukuda}, {Hanawa} \& {Sugitani}}{{Fukuda}
  et~al.}{2002}]{2002ApJ...568L.127F}
{Fukuda} N.,  {Hanawa} T.,    {Sugitani} K.,  2002, \apjl, 568, L127

\bibitem[\protect\citeauthoryear{{Fukui}, {Ohama}, {Hanaoka}, {Furukawa},
  {Torii}, {Dawson}, {Mizuno}, {Hasegawa}, {Fukuda} \& {Soga}}{{Fukui}
  et~al.}{2014}]{2014ApJ...780...36F}
{Fukui} Y.,  {Ohama} A.,  {Hanaoka} N.,  {Furukawa} N.,  {Torii} K.,  {Dawson}
  J.~R.,  {Mizuno} N.,  {Hasegawa} K.,  {Fukuda} T.,    {Soga} S.,  2014, \apj,
  780, 36

\bibitem[\protect\citeauthoryear{{Furukawa}, {Dawson}, {Ohama}, {Kawamura},
  {Mizuno}, {Onishi} \& {Fukui}}{{Furukawa} et~al.}{2009}]{2009ApJ...696L.115F}
{Furukawa} N.,  {Dawson} J.~R.,  {Ohama} A.,  {Kawamura} A.,  {Mizuno} N.,
  {Onishi} T.,    {Fukui} Y.,  2009, \apjl, 696, L115

\bibitem[\protect\citeauthoryear{{Gaczkowski}, {Preibisch}, {Ratzka},
  {Roccatagliata}, {Ohlendorf} \& {Zinnecker}}{{Gaczkowski}
  et~al.}{2013}]{2013A&A...549A..67G}
{Gaczkowski} B.,  {Preibisch} T.,  {Ratzka} T.,  {Roccatagliata} V.,
  {Ohlendorf} H.,    {Zinnecker} H.,  2013, \aap, 549, A67

\bibitem[\protect\citeauthoryear{{Garcia-Segura} \& {Franco}}{{Garcia-Segura}
  \& {Franco}}{1996}]{1996ApJ...469..171G}
{Garcia-Segura} G.,  {Franco} J.,  1996, \apj, 469, 171

\bibitem[\protect\citeauthoryear{{Getman}, {Feigelson}, {Garmire}, {Broos} \&
  {Wang}}{{Getman} et~al.}{2007}]{2007ApJ...654..316G}
{Getman} K.~V.,  {Feigelson} E.~D.,  {Garmire} G.,  {Broos} P.,    {Wang} J.,
  2007, \apj, 654, 316

\bibitem[\protect\citeauthoryear{{Getman}, {Feigelson}, {Sicilia-Aguilar},
  {Broos}, {Kuhn} \& {Garmire}}{{Getman} et~al.}{2012}]{2012MNRAS.426.2917G}
{Getman} K.~V.,  {Feigelson} E.~D.,  {Sicilia-Aguilar} A.,  {Broos} P.~S.,
  {Kuhn} M.~A.,    {Garmire} G.~P.,  2012, \mnras, 426, 2917

\bibitem[\protect\citeauthoryear{{Gouliermis}, {Chu}, {Henning}, {Brandner},
  {Gruendl}, {Hennekemper} \& {Hormuth}}{{Gouliermis}
  et~al.}{2008}]{2008ApJ...688.1050G}
{Gouliermis} D.~A.,  {Chu} Y.-H.,  {Henning} T.,  {Brandner} W.,  {Gruendl}
  R.~A.,  {Hennekemper} E.,    {Hormuth} F.,  2008, \apj, 688, 1050

\bibitem[\protect\citeauthoryear{{Gritschneder}, {Burkert}, {Naab} \&
  {Walch}}{{Gritschneder} et~al.}{2010}]{2010ApJ...723..971G}
{Gritschneder} M.,  {Burkert} A.,  {Naab} T.,    {Walch} S.,  2010, \apj, 723,
  971

\bibitem[\protect\citeauthoryear{{Gritschneder}, {Naab}, {Burkert}, {Walch},
  {Heitsch} \& {Wetzstein}}{{Gritschneder} et~al.}{2009}]{2009MNRAS.393...21G}
{Gritschneder} M.,  {Naab} T.,  {Burkert} A.,  {Walch} S.,  {Heitsch} F.,
  {Wetzstein} M.,  2009, \mnras, 393, 21

\bibitem[\protect\citeauthoryear{{Hatano}, {Kadowaki}, {Nakajima}, {Tamura},
  {Nagata}, {Sugitani}, {Tanab{\'e}}, {Kato}, {Kurita}, {Nishiyama}, {Baba},
  {Ishihara} \& {Sato}}{{Hatano} et~al.}{2006}]{2006AJ....132.2653H}
{Hatano} H.,  {Kadowaki} R.,  {Nakajima} Y.,  {Tamura} M.,  {Nagata} T.,
  {Sugitani} K.,  {Tanab{\'e}} T.,  {Kato} D.,  {Kurita} M.,  {Nishiyama} S.,
  {Baba} D.,  {Ishihara} A.,    {Sato} S.,  2006, \aj, 132, 2653

\bibitem[\protect\citeauthoryear{{Haworth} \& {Harries}}{{Haworth} \&
  {Harries}}{2012}]{2012MNRAS.420..562H}
{Haworth} T.~J.,  {Harries} T.~J.,  2012, \mnras, 420, 562

\bibitem[\protect\citeauthoryear{{Haworth}, {Harries} \& {Acreman}}{{Haworth}
  et~al.}{2012}]{2012MNRAS.426..203H}
{Haworth} T.~J.,  {Harries} T.~J.,    {Acreman} D.~M.,  2012, \mnras, 426, 203

\bibitem[\protect\citeauthoryear{{Haworth}, {Harries}, {Acreman} \&
  {Rundle}}{{Haworth} et~al.}{2013}]{2013MNRAS.431.3470H}
{Haworth} T.~J.,  {Harries} T.~J.,  {Acreman} D.~M.,    {Rundle} D.~A.,  2013,
  \mnras, 431, 3470

\bibitem[\protect\citeauthoryear{{Hosokawa} \& {Inutsuka}}{{Hosokawa} \&
  {Inutsuka}}{2005}]{2005ApJ...623..917H}
{Hosokawa} T.,  {Inutsuka} S.-i.,  2005, \apj, 623, 917

\bibitem[\protect\citeauthoryear{{Iwasaki}, {Inutsuka} \& {Tsuribe}}{{Iwasaki}
  et~al.}{2011a}]{2011ApJ...733...16I}
{Iwasaki} K.,  {Inutsuka} S.-i.,    {Tsuribe} T.,  2011a, \apj, 733, 16

\bibitem[\protect\citeauthoryear{{Iwasaki}, {Inutsuka} \& {Tsuribe}}{{Iwasaki}
  et~al.}{2011b}]{2011ApJ...733...17I}
{Iwasaki} K.,  {Inutsuka} S.-i.,    {Tsuribe} T.,  2011b, \apj, 733, 17

\bibitem[\protect\citeauthoryear{{Jiang}, {Yao}, {Yang}, {Ando}, {Kato},
  {Kawai}, {Kurita}, {Nagata}, {Nagayama}, {Nakajima}, {Nagashima}, {Sato},
  {Tamura}, {Nakaya} \& {Sugitani}}{{Jiang} et~al.}{2002}]{2002ApJ...577..245J}
{Jiang} Z.,  {Yao} Y.,  {Yang} J.,  {Ando} M.,  {Kato} D.,  {Kawai} T.,
  {Kurita} M.,  {Nagata} T.,  {Nagayama} T.,  {Nakajima} Y.,  {Nagashima} C.,
  {Sato} S.,  {Tamura} M.,  {Nakaya} H.,    {Sugitani} K.,  2002, \apj, 577,
  245

\bibitem[\protect\citeauthoryear{{Karr} \& {Martin}}{{Karr} \&
  {Martin}}{2003}]{2003ApJ...595..900K}
{Karr} J.~L.,  {Martin} P.~G.,  2003, \apj, 595, 900

\bibitem[\protect\citeauthoryear{{Kendrew}, {Simpson}, {Bressert}, {Povich},
  {Sherman}, {Lintott}, {Robitaille}, {Schawinski} \& {Wolf-Chase}}{{Kendrew}
  et~al.}{2012}]{2012ApJ...755...71K}
{Kendrew} S.,  {Simpson} R.,  {Bressert} E.,  {Povich} M.~S.,  {Sherman} R.,
  {Lintott} C.~J.,  {Robitaille} T.~P.,  {Schawinski} K.,    {Wolf-Chase} G.,
  2012, \apj, 755, 71

\bibitem[\protect\citeauthoryear{{Kennicutt} \& {Evans}}{{Kennicutt} \&
  {Evans}}{2012}]{2012ARA&A..50..531K}
{Kennicutt} R.~C.,  {Evans} N.~J.,  2012, \araa, 50, 531

\bibitem[\protect\citeauthoryear{{Klein}, {McKee} \& {Colella}}{{Klein}
  et~al.}{1994}]{1994ApJ...420..213K}
{Klein} R.~I.,  {McKee} C.~F.,    {Colella} P.,  1994, \apj, 420, 213

\bibitem[\protect\citeauthoryear{{Koenig}, {Allen}, {Gutermuth}, {Hora},
  {Brunt} \& {Muzerolle}}{{Koenig} et~al.}{2008}]{2008ApJ...688.1142K}
{Koenig} X.~P.,  {Allen} L.~E.,  {Gutermuth} R.~A.,  {Hora} J.~L.,  {Brunt}
  C.~M.,    {Muzerolle} J.,  2008, \apj, 688, 1142

\bibitem[\protect\citeauthoryear{{Koo}, {McKee}, {Lee}, {Lee}, {Lee}, {Moon},
  {Hong}, {Kaneda} \& {Onaka}}{{Koo} et~al.}{2008}]{2008ApJ...673L.147K}
{Koo} B.-C.,  {McKee} C.~F.,  {Lee} J.-J.,  {Lee} H.-G.,  {Lee} J.-E.,  {Moon}
  D.-S.,  {Hong} S.~S.,  {Kaneda} H.,    {Onaka} T.,  2008, \apjl, 673, L147

\bibitem[\protect\citeauthoryear{{Krumholz}, {Klein} \& {McKee}}{{Krumholz}
  et~al.}{2007}]{2007ApJ...656..959K}
{Krumholz} M.~R.,  {Klein} R.~I.,    {McKee} C.~F.,  2007, \apj, 656, 959

\bibitem[\protect\citeauthoryear{{Krumholz}, {Klein} \& {McKee}}{{Krumholz}
  et~al.}{2011}]{2011ApJ...740...74K}
{Krumholz} M.~R.,  {Klein} R.~I.,    {McKee} C.~F.,  2011, \apj, 740, 74

\bibitem[\protect\citeauthoryear{{Lee} \& {Chen}}{{Lee} \&
  {Chen}}{2007}]{2007ApJ...657..884L}
{Lee} H.-T.,  {Chen} W.~P.,  2007, \apj, 657, 884

\bibitem[\protect\citeauthoryear{{Lee}, {Chen}, {Zhang} \& {Hu}}{{Lee}
  et~al.}{2005}]{2005ApJ...624..808L}
{Lee} H.-T.,  {Chen} W.~P.,  {Zhang} Z.-W.,    {Hu} J.-Y.,  2005, \apj, 624,
  808

\bibitem[\protect\citeauthoryear{{Lefloch} \& {Lazareff}}{{Lefloch} \&
  {Lazareff}}{1994}]{1994A&A...289..559L}
{Lefloch} B.,  {Lazareff} B.,  1994, \aap, 289, 559

\bibitem[\protect\citeauthoryear{{Li} \& {Nakamura}}{{Li} \&
  {Nakamura}}{2006}]{2006ApJ...640L.187L}
{Li} Z.-Y.,  {Nakamura} F.,  2006, \apjl, 640, L187

\bibitem[\protect\citeauthoryear{{Liu}, {Wu}, {Li}, {Yuan}, {Liu} \&
  {Dong}}{{Liu} et~al.}{2015}]{2015ApJ...798...30L}
{Liu} H.-L.,  {Wu} Y.,  {Li} J.,  {Yuan} J.-H.,  {Liu} T.,    {Dong} X.,  2015,
  \apj, 798, 30

\bibitem[\protect\citeauthoryear{{Liu}, {Wu}, {Zhang} \& {Qin}}{{Liu}
  et~al.}{2012}]{2012ApJ...751...68L}
{Liu} T.,  {Wu} Y.,  {Zhang} H.,    {Qin} S.-L.,  2012, \apj, 751, 68

\bibitem[\protect\citeauthoryear{{Matsuyanagi}, {Itoh}, {Sugitani}, {Oasa},
  {Mukai} \& {Tamura}}{{Matsuyanagi} et~al.}{2006}]{2006PASJ...58L..29M}
{Matsuyanagi} I.,  {Itoh} Y.,  {Sugitani} K.,  {Oasa} Y.,  {Mukai} T.,
  {Tamura} M.,  2006, \pasj, 58, L29

\bibitem[\protect\citeauthoryear{{Matzner}}{{Matzner}}{2002}]{2002ApJ...566..302M}
{Matzner} C.~D.,  2002, \apj, 566, 302

\bibitem[\protect\citeauthoryear{{Minier}, {Andr{\'e}}, {Bergman}, {Motte},
  {Wyrowski} \& {Le Pennec}}{{Minier} et~al.}{2009}]{2009A&A...501L...1M}
{Minier} V.,  {Andr{\'e}} P.,  {Bergman} P.,  {Motte} F.,  {Wyrowski} F.,
  {Le Pennec} J.,  2009, \aap, 501, L1

\bibitem[\protect\citeauthoryear{{Morgan}, {Thompson}, {Urquhart}, {White} \&
  {Miao}}{{Morgan} et~al.}{2004}]{2004A&A...426..535M}
{Morgan} L.~K.,  {Thompson} M.~A.,  {Urquhart} J.~S.,  {White} G.~J.,    {Miao}
  J.,  2004, \aap, 426, 535

\bibitem[\protect\citeauthoryear{{Myers}, {Klein}, {Krumholz} \&
  {McKee}}{{Myers} et~al.}{2014}]{2014MNRAS.tmp..421M}
{Myers} A.~T.,  {Klein} R.~I.,  {Krumholz} M.~R.,    {McKee} C.~F.,  2014,
  \mnras

\bibitem[\protect\citeauthoryear{{Nakajima}, {Kato}, {Nagata}, {Tamura},
  {Sato}, {Sugitani}, {Nagashima}, {Nagayama}, {Iwata}, {Ita}, {Tanabe},
  {Kurita}, {Nakaya} \& {Baba}}{{Nakajima} et~al.}{2005}]{2005AJ....129..776N}
{Nakajima} Y.,  {Kato} D.,  {Nagata} T.,  {Tamura} M.,  {Sato} S.,  {Sugitani}
  K.,  {Nagashima} C.,  {Nagayama} T.,  {Iwata} I.,  {Ita} Y.,  {Tanabe} T.,
  {Kurita} M.,  {Nakaya} H.,    {Baba} D.,  2005, \aj, 129, 776

\bibitem[\protect\citeauthoryear{{Negueruela}, {Marco}, {Israel} \&
  {Bernabeu}}{{Negueruela} et~al.}{2007}]{2007A&A...471..485N}
{Negueruela} I.,  {Marco} A.,  {Israel} G.~L.,    {Bernabeu} G.,  2007, \aap,
  471, 485

\bibitem[\protect\citeauthoryear{{Oey}, {Watson}, {Kern} \& {Walth}}{{Oey}
  et~al.}{2005}]{2005AJ....129..393O}
{Oey} M.~S.,  {Watson} A.~M.,  {Kern} K.,    {Walth} G.~L.,  2005, \aj, 129,
  393

\bibitem[\protect\citeauthoryear{{Ohama}, {Dawson}, {Furukawa}, {Kawamura},
  {Moribe}, {Yamamoto}, {Okuda}, {Mizuno}, {Onishi}, {Maezawa}, {Minamidani},
  {Mizuno} \& {Fukui}}{{Ohama} et~al.}{2010}]{2010ApJ...709..975O}
{Ohama} A.,  {Dawson} J.~R.,  {Furukawa} N.,  {Kawamura} A.,  {Moribe} N.,
  {Yamamoto} H.,  {Okuda} T.,  {Mizuno} N.,  {Onishi} T.,  {Maezawa} H.,
  {Minamidani} T.,  {Mizuno} A.,    {Fukui} Y.,  2010, \apj, 709, 975

\bibitem[\protect\citeauthoryear{{Paron}, {Petriella} \& {Ortega}}{{Paron}
  et~al.}{2011}]{2011A&A...525A.132P}
{Paron} S.,  {Petriella} A.,    {Ortega} M.~E.,  2011, \aap, 525, A132

\bibitem[\protect\citeauthoryear{{Peng}, {Wyrowski}, {van der Tak}, {Menten} \&
  {Walmsley}}{{Peng} et~al.}{2010}]{2010A&A...520A..84P}
{Peng} T.-C.,  {Wyrowski} F.,  {van der Tak} F.~F.~S.,  {Menten} K.~M.,
  {Walmsley} C.~M.,  2010, \aap, 520, A84

\bibitem[\protect\citeauthoryear{{Peters}, {Banerjee}, {Klessen}, {Mac Low},
  {Galv{\'a}n-Madrid} \& {Keto}}{{Peters} et~al.}{2010}]{2010ApJ...711.1017P}
{Peters} T.,  {Banerjee} R.,  {Klessen} R.~S.,  {Mac Low} M.,
  {Galv{\'a}n-Madrid} R.,    {Keto} E.~R.,  2010, \apj, 711, 1017

\bibitem[\protect\citeauthoryear{{Peters}, {Banerjee}, {Klessen} \& {Mac
  Low}}{{Peters} et~al.}{2011}]{2011ApJ...729...72P}
{Peters} T.,  {Banerjee} R.,  {Klessen} R.~S.,    {Mac Low} M.-M.,  2011, \apj,
  729, 72

\bibitem[\protect\citeauthoryear{{Preibisch}, {Ratzka}, {Kuderna}, {Ohlendorf},
  {King}, {Hodgkin}, {Irwin}, {Lewis}, {McCaughrean} \&
  {Zinnecker}}{{Preibisch} et~al.}{2011}]{2011A&A...530A..34P}
{Preibisch} T.,  {Ratzka} T.,  {Kuderna} B.,  {Ohlendorf} H.,  {King} R.~R.,
  {Hodgkin} S.,  {Irwin} M.,  {Lewis} J.~R.,  {McCaughrean} M.~J.,
  {Zinnecker} H.,  2011, \aap, 530, A34

\bibitem[\protect\citeauthoryear{{Preibisch} \& {Zinnecker}}{{Preibisch} \&
  {Zinnecker}}{2001}]{2001ASPC..243..791P}
{Preibisch} T.,  {Zinnecker} H.,  2001, in {Montmerle} T.,  {Andr{\'e}} P.,
  eds, From Darkness to Light: Origin and Evolution of Young Stellar Clusters
  Vol.~243 of Astronomical Society of the Pacific Conference Series, {Triggered
  Star Formation in the Scorpius-Centaurus OB Association (Sco OB2)}.
p.~791

\bibitem[\protect\citeauthoryear{{Puga}, {Hony}, {Neiner}, {Lenorzer},
  {Hubert}, {Waters}, {Cusano} \& {Ripepi}}{{Puga}
  et~al.}{2009}]{2009A&A...503..107P}
{Puga} E.,  {Hony} S.,  {Neiner} C.,  {Lenorzer} A.,  {Hubert} A.,  {Waters}
  L.~B.~F.~M.,  {Cusano} F.,    {Ripepi} V.,  2009, \aap, 503, 107

\bibitem[\protect\citeauthoryear{{Roman-Lopes}}{{Roman-Lopes}}{2009}]{2009MNRAS.398.1368R}
{Roman-Lopes} A.,  2009, \mnras, 398, 1368

\bibitem[\protect\citeauthoryear{{Samal}, {Zavagno}, {Deharveng}, {Molinari},
  {Ojha}, {Paradis}, {Tig{\'e}}, {Pandey} \& {Russeil}}{{Samal}
  et~al.}{2014}]{2014A&A...566A.122S}
{Samal} M.~R.,  {Zavagno} A.,  {Deharveng} L.,  {Molinari} S.,  {Ojha} D.~K.,
  {Paradis} D.,  {Tig{\'e}} J.,  {Pandey} A.~K.,    {Russeil} D.,  2014, \aap,
  566, A122

\bibitem[\protect\citeauthoryear{{Shimajiri}, {Takahashi}, {Takakuwa}, {Saito}
  \& {Kawabe}}{{Shimajiri} et~al.}{2008}]{2008ApJ...683..255S}
{Shimajiri} Y.,  {Takahashi} S.,  {Takakuwa} S.,  {Saito} M.,    {Kawabe} R.,
  2008, \apj, 683, 255

\bibitem[\protect\citeauthoryear{{Shore}}{{Shore}}{1981}]{1981ApJ...249...93S}
{Shore} S.~N.,  1981, \apj, 249, 93

\bibitem[\protect\citeauthoryear{{Smith}, {Egan}, {Carey}, {Price}, {Morse} \&
  {Price}}{{Smith} et~al.}{2000}]{2000ApJ...532L.145S}
{Smith} N.,  {Egan} M.~P.,  {Carey} S.,  {Price} S.~D.,  {Morse} J.~A.,
  {Price} P.~A.,  2000, \apjl, 532, L145

\bibitem[\protect\citeauthoryear{{Smith}, {Stassun} \& {Bally}}{{Smith}
  et~al.}{2005}]{2005AJ....129..888S}
{Smith} N.,  {Stassun} K.~G.,    {Bally} J.,  2005, \aj, 129, 888

\bibitem[\protect\citeauthoryear{{Snider}, {Hester}, {Desch}, {Healy} \&
  {Bally}}{{Snider} et~al.}{2009}]{2009ApJ...700..506S}
{Snider} K.~D.,  {Hester} J.~J.,  {Desch} S.~J.,  {Healy} K.~R.,    {Bally} J.,
   2009, \apj, 700, 506

\bibitem[\protect\citeauthoryear{{Solomon}, {Sanders} \& {Scoville}}{{Solomon}
  et~al.}{1979}]{1979IAUS...84...35S}
{Solomon} P.~M.,  {Sanders} D.~B.,    {Scoville} N.~Z.,  1979, in {Burton}
  W.~B.,  ed., The Large-Scale Characteristics of the Galaxy Vol.~84 of IAU
  Symposium, {Giant molecular clouds in the Galaxy - Distribution, mass, size
  and age}.
pp 35--52

\bibitem[\protect\citeauthoryear{{Stanke}, {Smith}, {Gredel} \&
  {Szokoly}}{{Stanke} et~al.}{2002}]{2002A&A...393..251S}
{Stanke} T.,  {Smith} M.~D.,  {Gredel} R.,    {Szokoly} G.,  2002, \aap, 393,
  251

\bibitem[\protect\citeauthoryear{{Sugitani}, {Tamura} \& {Ogura}}{{Sugitani}
  et~al.}{1995}]{1995ApJ...455L..39S}
{Sugitani} K.,  {Tamura} M.,    {Ogura} K.,  1995, \apjl, 455, L39

\bibitem[\protect\citeauthoryear{{Tackenberg}, {Beuther}, {Plume}, {Henning},
  {Stil}, {Walmsley}, {Schuller} \& {Schmiedeke}}{{Tackenberg}
  et~al.}{2013}]{2013A&A...550A.116T}
{Tackenberg} J.,  {Beuther} H.,  {Plume} R.,  {Henning} T.,  {Stil} J.,
  {Walmsley} M.,  {Schuller} F.,    {Schmiedeke} A.,  2013, \aap, 550, A116

\bibitem[\protect\citeauthoryear{{Tan}}{{Tan}}{2000}]{2000ApJ...536..173T}
{Tan} J.~C.,  2000, \apj, 536, 173

\bibitem[\protect\citeauthoryear{{Tenorio-Tagle} \&
  {Bodenheimer}}{{Tenorio-Tagle} \& {Bodenheimer}}{1988}]{1988ARA&A..26..145T}
{Tenorio-Tagle} G.,  {Bodenheimer} P.,  1988, \araa, 26, 145

\bibitem[\protect\citeauthoryear{{Thompson}, {Urquhart}, {Moore} \&
  {Morgan}}{{Thompson} et~al.}{2012}]{2012MNRAS.421..408T}
{Thompson} M.~A.,  {Urquhart} J.~S.,  {Moore} T.~J.~T.,    {Morgan} L.~K.,
  2012, \mnras, 421, 408

\bibitem[\protect\citeauthoryear{{Thompson}, {White}, {Morgan}, {Miao},
  {Fridlund} \& {Huldtgren-White}}{{Thompson}
  et~al.}{2004}]{2004A&A...414.1017T}
{Thompson} M.~A.,  {White} G.~J.,  {Morgan} L.~K.,  {Miao} J.,  {Fridlund}
  C.~V.~M.,    {Huldtgren-White} M.,  2004, \aap, 414, 1017

\bibitem[\protect\citeauthoryear{{Torii}, {Enokiya}, {Sano}, {Yoshiike},
  {Hanaoka}, {Ohama}, {Furukawa}, {Dawson}, {Moribe}, {Oishi}, {Nakashima},
  {Okuda}, {Yamamoto}, {Kawamura}, {Mizuno}, {Maezawa}, {Onishi}, {Mizuno} \&
  {Fukui}}{{Torii} et~al.}{2011}]{2011ApJ...738...46T}
{Torii} K.,  {Enokiya} R.,  {Sano} H.,  {Yoshiike} S.,  {Hanaoka} N.,  {Ohama}
  A.,  {Furukawa} N.,  {Dawson} J.~R.,  {Moribe} N.,  {Oishi} K.,  {Nakashima}
  Y.,  {Okuda} T.,  {Yamamoto} H.,  {Kawamura} A.,  {Mizuno} N.,  {Maezawa} H.,
   {Onishi} T.,  {Mizuno} A.,    {Fukui} Y.,  2011, \apj, 738, 46

\bibitem[\protect\citeauthoryear{{Tremblin}, {Audit}, {Minier}, {Schmidt} \&
  {Schneider}}{{Tremblin} et~al.}{2012}]{2012A&A...546A..33T}
{Tremblin} P.,  {Audit} E.,  {Minier} V.,  {Schmidt} W.,    {Schneider} N.,
  2012, \aap, 546, A33

\bibitem[\protect\citeauthoryear{{Tremblin}, {Audit}, {Minier} \&
  {Schneider}}{{Tremblin} et~al.}{2012}]{2012A&A...538A..31T}
{Tremblin} P.,  {Audit} E.,  {Minier} V.,    {Schneider} N.,  2012, \aap, 538,
  A31

\bibitem[\protect\citeauthoryear{{Urquhart}, {Morgan} \& {Thompson}}{{Urquhart}
  et~al.}{2009}]{2009A&A...497..789U}
{Urquhart} J.~S.,  {Morgan} L.~K.,    {Thompson} M.~A.,  2009, \aap, 497, 789

\bibitem[\protect\citeauthoryear{{Urquhart}, {Thompson}, {Morgan},
  {Pestalozzi}, {White} \& {Muna}}{{Urquhart}
  et~al.}{2007}]{2007A&A...467.1125U}
{Urquhart} J.~S.,  {Thompson} M.~A.,  {Morgan} L.~K.,  {Pestalozzi} M.~R.,
  {White} G.~J.,    {Muna} D.~N.,  2007, \aap, 467, 1125

\bibitem[\protect\citeauthoryear{{Walborn}, {Barb{\'a}}, {Brandner}, {Rubio},
  {Grebel} \& {Probst}}{{Walborn} et~al.}{1999}]{1999AJ....117..225W}
{Walborn} N.~R.,  {Barb{\'a}} R.~H.,  {Brandner} W.,  {Rubio} M.,  {Grebel}
  E.~K.,    {Probst} R.~G.,  1999, \aj, 117, 225

\bibitem[\protect\citeauthoryear{{Walborn}, {Ma{\'{\i}}z-Apell{\'a}niz} \&
  {Barb{\'a}}}{{Walborn} et~al.}{2002}]{2002AJ....124.1601W}
{Walborn} N.~R.,  {Ma{\'{\i}}z-Apell{\'a}niz} J.,    {Barb{\'a}} R.~H.,  2002,
  \aj, 124, 1601

\bibitem[\protect\citeauthoryear{{Walch}, {Whitworth}, {Bisbas}, {W{\"u}nsch}
  \& {Hubber}}{{Walch} et~al.}{2013}]{2013MNRAS.435..917W}
{Walch} S.,  {Whitworth} A.~P.,  {Bisbas} T.~G.,  {W{\"u}nsch} R.,    {Hubber}
  D.~A.,  2013, \mnras, 435, 917

\bibitem[\protect\citeauthoryear{{Watson}, {Hanspal} \& {Mengistu}}{{Watson}
  et~al.}{2010}]{2010ApJ...716.1478W}
{Watson} C.,  {Hanspal} U.,    {Mengistu} A.,  2010, \apj, 716, 1478

\bibitem[\protect\citeauthoryear{{Whitworth}, {Bhattal}, {Chapman}, {Disney} \&
  {Turner}}{{Whitworth} et~al.}{1994}]{1994A&A...290..421W}
{Whitworth} A.~P.,  {Bhattal} A.~S.,  {Chapman} S.~J.,  {Disney} M.~J.,
  {Turner} J.~A.,  1994, \aap, 290, 421

\bibitem[\protect\citeauthoryear{{Wilking}, {Harvey}, {Lada}, {Joy} \&
  {Doering}}{{Wilking} et~al.}{1984}]{1984ApJ...279..291W}
{Wilking} B.~A.,  {Harvey} P.~M.,  {Lada} C.~J.,  {Joy} M.,    {Doering} C.~R.,
   1984, \apj, 279, 291

\bibitem[\protect\citeauthoryear{{Williams}, {Ward-Thompson} \&
  {Whitworth}}{{Williams} et~al.}{2001}]{2001MNRAS.327..788W}
{Williams} R.~J.~R.,  {Ward-Thompson} D.,    {Whitworth} A.~P.,  2001, \mnras,
  327, 788

\bibitem[\protect\citeauthoryear{{W{\"u}nsch}, {Dale}, {Palou{\v s}} \&
  {Whitworth}}{{W{\"u}nsch} et~al.}{2010}]{2010MNRAS.407.1963W}
{W{\"u}nsch} R.,  {Dale} J.~E.,  {Palou{\v s}} J.,    {Whitworth} A.~P.,  2010,
  \mnras, 407, 1963

\bibitem[\protect\citeauthoryear{{W{\"u}nsch} \& {Palou{\v s}}}{{W{\"u}nsch} \&
  {Palou{\v s}}}{2001}]{2001A&A...374..746W}
{W{\"u}nsch} R.,  {Palou{\v s}} J.,  2001, \aap, 374, 746

\bibitem[\protect\citeauthoryear{{Xu} \& {Wang}}{{Xu} \&
  {Wang}}{2012}]{2012A&A...543A..24X}
{Xu} J.-L.,  {Wang} J.-J.,  2012, \aap, 543, A24

\bibitem[\protect\citeauthoryear{{Xu}, {Wang} \& {Liu}}{{Xu}
  et~al.}{2013}]{2013A&A...559A.113X}
{Xu} J.-L.,  {Wang} J.-J.,    {Liu} X.-L.,  2013, \aap, 559, A113

\bibitem[\protect\citeauthoryear{{Yuan}, {Wu}, {Li}, {Yu} \& {Miller}}{{Yuan}
  et~al.}{2013}]{2013MNRAS.429..954Y}
{Yuan} J.-H.,  {Wu} Y.,  {Li} J.~Z.,  {Yu} W.,    {Miller} M.,  2013, \mnras,
  429, 954

\bibitem[\protect\citeauthoryear{{Zavagno}, {Anderson}, {Russeil}, {Morgan},
  {Stringfellow}, {Deharveng}, {Rod{\'o}n}, {Robitaille} \&
  {Mottram}}{{Zavagno} et~al.}{2010}]{2010A&A...518L.101Z}
{Zavagno} A.,  {Anderson} L.~D.,  {Russeil} D.,  {Morgan} L.,  {Stringfellow}
  G.~S.,  {Deharveng} L.,  {Rod{\'o}n} J.~A.,  {Robitaille} T.~P.,    {Mottram}
  J.~C.,  2010, \aap, 518, L101

\bibitem[\protect\citeauthoryear{{Zavagno}, {Deharveng}, {Comer{\'o}n},
  {Brand}, {Massi}, {Caplan} \& {Russeil}}{{Zavagno}
  et~al.}{2006}]{2006A&A...446..171Z}
{Zavagno} A.,  {Deharveng} L.,  {Comer{\'o}n} F.,  {Brand} J.,  {Massi} F.,
  {Caplan} J.,    {Russeil} D.,  2006, \aap, 446, 171

\bibitem[\protect\citeauthoryear{{Zavagno}, {Pomar{\`e}s}, {Deharveng},
  {Hosokawa}, {Russeil} \& {Caplan}}{{Zavagno}
  et~al.}{2007}]{2007A&A...472..835Z}
{Zavagno} A.,  {Pomar{\`e}s} M.,  {Deharveng} L.,  {Hosokawa} T.,  {Russeil}
  D.,    {Caplan} J.,  2007, \aap, 472, 835

\bibitem[\protect\citeauthoryear{{Zavagno}, {Russeil}, {Motte}, {Anderson},
  {Deharveng}, {Rod{\'o}n}, {Bontemps} \& {Abergel}}{{Zavagno}
  et~al.}{2010}]{2010A&A...518L..81Z}
{Zavagno} A.,  {Russeil} D.,  {Motte} F.,  {Anderson} L.~D.,  {Deharveng} L.,
  {Rod{\'o}n} J.~A.,  {Bontemps} S.,    {Abergel} A.,  2010, \aap, 518, L81

\end{thebibliography}
\appendix
\section{Observational work}
Observational papers referred to are listed below, together with their triggering signposts in a checkbox format.\\
\onecolumn
\begin{longtable}{rrrrrrrrrl}
&&Shell/&&Pil./&Rel./Dyn.&Age&Elong.&Gas\\
Reference&Target&HIIR/IF&BRC&Com.&ages&grad.&clus.&Int.&Driver\\
\hline
{\cite{2009AJ....138..975B}}&Sh 254&-&-&-&X&X&-&-&HIIR\\
{\cite{2010ApJ...713..883B}}&RCW34&X&X&-&-&-&-&-&HIIR\\
{\cite{2010ApJ...712..797B}}&Vul OB1&X&-&X&-&-&-&-&SNR\\
{\cite{2011MNRAS.415.1202C}}&W5 E&-&X&-&-&X&X&-&HIIR\\
{\cite{2009MNRAS.396..964C}}&Various BRCs&X&X&-&X&X&X&-&HIIR\\
{\cite{2010ApJ...717.1067C}}&SFO 38&-&X&-&X&X&X&-&HIIR\\
{\cite{2007ApJ...670..428C}}&Many bubbles&X&-&-&-&-&-&-&Wind\\
{\cite{2006ApJ...649..759C}}&Many bubbles&X&-&-&-&-&-&-&Wind\\
{\cite{2014MNRAS.438.1089C}}&G126.1-0.8-1.4&X&-&-&X&-&-&-&Wind/SN\\
{\cite{2004A&A...427..839C}}&WR48A&X&-&-&X&-&-&X&WR wind\\
{\cite{2010A&A...523A...6D}}&Many bubbles&X&-&X&-&-&-&X&HIIR\\
{\cite{2009A&A...496..177D}}&RCW120&X&-&-&-&-&-&X&HIIR\\
{\cite{2008A&A...482..585D}}&Sh2--212&X&-&-&X&-&-&X&HIIR\\
{\cite{2006A&A...458..191D}}&Sh2--219&X&-&-&-&-&-&X&HIIR\\
{\cite{2005A&A...433..565D}}&Many bubbles&X&-&-&-&-&-&X&HIIR\\
{\cite{2003A&A...408L..25D}}&Sh 104&X&-&-&-&-&-&-&HIIR\\
{\cite{2003A&A...399.1135D}}&Sh 217/219&X&-&-&-&X&-&X&HIIR\\
{\cite{2012ApJ...756..151D}}&G8.14&X&-&-&X&-&-&X&HIIR\\
{\cite{2014MNRAS.444..376E}}&IC 2574&X&-&-&X&-&-&-&Supershell\\
{\cite{2002ApJ...568L.127F}}&M16&-&-&-&X&X&X&-&HIIR\\
{\cite{2013A&A...549A..67G}}&Carina&X&-&X&-&-&-&-&HIIR/wind\\
{\cite{2012MNRAS.426.2917G}}&IC1396&-&-&-&-&X&-&-&HIIR\\
{\cite{2007ApJ...654..316G}}&IC1396&X&X&-&-&X&X&-&HIIR\\
{\cite{2008ApJ...688.1050G}}&NGC346&X&-&-&-&-&-&-&Wind/SN\\
{\cite{2006AJ....132.2653H}}&LH9/N11&X&-&-&X&-&-&X&Wind/SN\\
{\cite{2002ApJ...577..245J}}&M17&X&-&-&X&-&-&-&HIIR\\
{\cite{2003ApJ...595..900K}}&W5&X&X&-&X&-&-&X&HIIR\\
{\cite{2012ApJ...755...71K}}&Many bubbles&X&-&-&-&-&-&-&HIIR\\
{\cite{2008ApJ...688.1142K}}&W5&X&X&X&X&-&X&-&HIIR\\
{\cite{2008ApJ...673L.147K}}&G54.1&X&-&-&-&-&-&-&Wind/SNR\\
{\cite{2007ApJ...657..884L}}&Various&-&X&-&-&X&-&-&HIIR\\
{\cite{2005ApJ...624..808L}}&Orion BRCs&X&X&-&X&X&X&-&HIIR\\
{\cite{2012ApJ...751...68L}}&HD211853&X&-&-&-&X&-&-&WR star\\
{\cite{2015ApJ...798...30L}}&G24.136+00.436&X&-&-&X&X&-&X&HIIR\\
{\cite{2006PASJ...58L..29M}}&BRC14&-&X&-&-&X&X&-&HIIR\\
{\cite{2009A&A...501L...1M}}&G327&X&-&-&-&-&-&-&HIIR\\
{\cite{2004A&A...426..535M}}&Various BRCs&-&X&-&-&-&-&X&HIIR\\
{\cite{2005AJ....129..776N}}&N159/160&X&-&-&-&X&-&-&HIIR\\
{\cite{2007A&A...471..485N}}&NGC1893&X&X&X&X&-&-&-&HIIR\\
{\cite{2005AJ....129..393O}}&W4/5&X&-&-&X&-&-&-&HIIR\\
{\cite{2011A&A...525A.132P}}&G35.673&X&-&-&-&-&-&X&HIIR\\
{\cite{2010A&A...520A..84P}}&W49A&X&-&-&X&-&-&-&Wind\\
{\cite{2011A&A...530A..34P}}&Carina&X&-&X&-&-&-&-&HIIR/wind\\
{\cite{2009A&A...503..107P}}&Sh2--284&X&X&X&-&-&-&-&HIIR\\
{\cite{2009MNRAS.398.1368R}}&Sh2--307&X&-&-&-&-&-&-&Wind\\
{\cite{2014A&A...566A.122S}}&Sh2--90&X&-&-&X&-&-&X&HIIR\\
{\cite{2008ApJ...683..255S}}&OMC--2/3&-&-&-&X&-&-&X&Outflows\\
{\cite{2005AJ....129..888S}}&Carina&X&-&X&-&-&-&-&HIIR/wind\\
{\cite{2000ApJ...532L.145S}}&Carina&X&X&X&-&-&-&-&HIIR/wind\\
{\cite{2009ApJ...700..506S}}&NGC2467&X&-&-&X&X&-&-&HIIR\\
{\cite{2002A&A...393..251S}}&Orion&X&-&X&-&X&-&-&HIIR/wind\\
{\cite{1995ApJ...455L..39S}}&Various BRCs&X&X&-&-&X&X&-&HIIR\\
{\cite{2012MNRAS.421..408T}}&Many bubbles&X&-&-&-&-&-&-&HIIR\\
{\cite{2004A&A...414.1017T}}&IC1848&-&X&X&X&-&-&X&HIIR\\
{\cite{2009A&A...497..789U}}&Various BRCs&X&X&-&-&-&-&X&HIIR\\
{\cite{2007A&A...467.1125U}}&SFO75&-&X&-&X&-&-&X&HIIR\\
{\cite{2002AJ....124.1601W}}&30 Dor&X&-&X&-&-&-&-&HIIR/wind\\
{\cite{1999AJ....117..225W}}&30 Dor&X&-&-&X&-&-&-&HIIR/wind\\
{\cite{2010ApJ...716.1478W}}&Many bubbles&X&-&-&-&-&-&-&HIIR\\
{\cite{1984ApJ...279..291W}}&W5&X&-&-&X&-&-&X&HIIR\\
&&Shell/&&Pil./&Rel./Dyn.&Age&Elong.&Gas\\
Reference&Target&HIIR/IF&BRC&Com.&ages&grad.&clus.&Int.&Driver\\
\hline
{\cite{2013A&A...559A.113X}}&G38.91&X&-&-&X&-&-&-&HIIR\\
{\cite{2012A&A...543A..24X}}&NGC6823&-&-&-&X&-&-&-&SNR\\
{\cite{2013MNRAS.429..954Y}}&L1174&X&-&-&-&X&-&-&Ae/Be wind\\
{\cite{2010A&A...518L.101Z}}&N49&X&-&-&-&-&-&-&HIIR\\
{\cite{2010A&A...518L..81Z}}&RCW120&X&-&-&-&-&-&-&HIIR\\
{\cite{2007A&A...472..835Z}}&RCW120&X&-&-&-&-&-&-&HIIR\\
{\cite{2006A&A...446..171Z}}&RCW79&X&-&-&-&-&-&-&HIIR\\
\hline
Totals &&55&18&12&27&18&9&19&67\\
\caption{A selection of observational papers on feedback characterised by the markers or indicators used. Key to abbreviations: Shell -- proximity to shell, HII region etc.; BRC -- proximity to a bright--rimmed cloud; Pil./Com. -- proximity to a pillar/cometary cloud; Rel/Dyn ages -- use of ages of stars relative to dynamical timescale of feedback--driven system; Age grad. -- use of stellar age gradient pointing from feedback source; Elong clus -- use of elongated shape of stellar system pointing to feedback source; Gas Int. -- evidence of strong interaction between the feedback source and the triggered region; Driver -- likely source of feedback.}
\label{tab:bigtable}
\end{longtable}
\twocolumn

\label{lastpage}

\end{document}